\documentclass{aa}

\usepackage{graphicx}
\usepackage{savesym}
\usepackage{amsmath}
\savesymbol{iint}
\usepackage{txfonts}
\restoresymbol{TXF}{iint}
\usepackage{rotating}
\usepackage{longtable,lscape}

\newcommand{\sub}[1]{\ensuremath{_\mathrm{#1}}}

\DeclareFontFamily{U}{euc}{}
\DeclareFontShape{U}{euc}{m}{n}{<-6>eurm5<6-8>eurm7<8->eurm10}{}%
\DeclareSymbolFont{AMSc}{U}{euc}{m}{n} 
\DeclareMathSymbol{\umu}{\mathord}{AMSc}{"16}

\defcitealias{bonavita14}{Paper II}

\begin{document}

   \title{SPOTS: The Search for Planets Orbiting Two Stars}

   \subtitle{I. Survey description and first observations%
\thanks{Based on data collected at the European Southern 
Observatory, Chile (ESO Programmes 088.C-0291 and 090.C-0416) 
and at the 
Subaru Telescope, which is operated by the National Astronomical 
Observatory of Japan.}}

   \author{C. Thalmann\inst{1}
          \and
          S. Desidera\inst{2}
          \and
          M. Bonavita\inst{2}
          \and
          M. Janson\inst{3}
          \and
          T. Usuda\inst{4,5}
          \and
          T. Henning\inst{6}
          \and
          R. K\"{o}hler\inst{6}
          \and
          J. Carson\inst{7,6}
          \and
          A. Boccaletti\inst{8}
          \and
          C. Bergfors\inst{9}
          \and
          W. Brandner\inst{6}
          \and
          M. Feldt\inst{6}
          \and
          M. Goto\inst{10}
          \and
          H. Klahr\inst{6}
          \and
          F. Marzari\inst{11}
          \and
          C. Mordasini\inst{6}
          }

   \institute{Institute for Astronomy, ETH Zurich, Wolfgang-Pauli-Strasse 27, 
   			8093 Zurich, Switzerland\\
            \email{thalmann@phys.ethz.ch}
            \and
            INAF -- Osservatorio Astronomico di Padova,  
              Vicolo dell'Osservatorio 5, I-35122, Padova, Italy
            \and
            Astrophysics Research Center, Queen's University Belfast, Belfast,
            Northern Ireland, UK
            \and
            National Astronomical Observatory of Japan, 2-21-1 Osawa, 
            Mitaka, 181-8588 Tokyo, Japan
            \and
            Department of Astronomical Sciences, Graduate University for 
            Advanced Studies (Sokendai), Mitaka, 181-8858 Tokyo, Japan
            \and
            Max Planck Institute for Astronomy, K\"onigstuhl 17, 
            69117 Heidelberg, Germany
            \and
            Department of Physics \& Astronomy, College of Charleston, 
            58 Coming Street, Charleston, SC 29424, USA
            \and
            Observatoire de Meudon, LESIA, bat 17, 5 pl.\ J.\ Janssen, 
            92195 Meudon, France
            \and
            Department of Physics \& Astronomy, University College London,
            Gower Street,
            London WC1E 6BT,
            UK
            \and
            Universit\"atssternwerte M\"unchen, Scheinerstr. 1, D-81679 
            Munich, Germany
            \and
            Dipartimento di Fisica, University of Padova, Via Marzolo 8, 
            35131 Padova, Italy
            }

   \date{Draft version}

 
  \abstract{Direct imaging surveys for exoplanets 
        commonly exclude binary stars from their target lists, 
        leaving a large part of the overall planet demography 
        unexplored. 
        To address this gap in our understanding of 
        planet formation and evolution, we have launched the 
        first direct imaging survey dedicated to 
        circumbinary planets: SPOTS, the Search for Planets Orbiting
        Two Stars.  In this paper, we discuss the theoretical 
        context, scientific merit,
        and technical feasibility of such observations, describe
        the target sample and observational strategy of our survey, 
        and report on the first results from our pilot survey of 26
        targets with the VLT NaCo facility.  While we have not found
        any confirmed substellar companions to date, a number of 
        promising candidate companions remain to be tested for 
        common proper motion in upcoming follow-up observations.
        We also report on the astrometry of the three resolved 
        binaries in our target sample.
        This pilot survey constitutes a successful proof of concept 
        for our survey strategy and paves the way for a second 
        stage of exploratory observations with VLT SPHERE.
    }


   \maketitle
%



\section{Introduction}

While binary star systems have traditionally been assumed to be hostile 
environments for the formation and survival of planets, a growing body of
theoretical and observational evidence demonstrates that this is not the
case.  Most exoplanet surveys based on the Doppler spectroscopy or 
direct imaging techniques to date have made active efforts to exclude 
at least some parameter ranges of known binaries from their target lists
\citep[e.g.,][]{jones02, wright04, marcy05}.
In particular, binary target 
stars present technical obstacles to Doppler spectroscopy,
which is responsible for most of the confirmed exoplanet discoveries
prior to 2014.
It may then come as a surprise that, in spite of these selection 
biases, 78 of the 729 exoplanets listed in the Exoplanets.org database
\citep{2011PASP..123..412W} by November 2013 reside in a multiple system,
indicating a large and mostly unexplored
population of such planets.  

Various theoretical and observational studies have been conducted on the
impact of binarity on the properties of planets 
\citep[see e.g.\ the book ``Planets in Binaries'',][]{2010ASSL..366.....H}.
Since most of these studies are based on Doppler spectroscopy 
(`radial velocity')
observations, they focus on planets on short-period circumstellar 
(s-type) orbits around 
components of wide binary systems.  The presence of a distant companion
star is expected to influence the planet formation process in such systems
through truncation and dynamical heating of the protoplanetary disk
\citep[e.g.,][]{artymowicz94}.  
The main result of these studies is that binarity may
change the shape of the planet demography, but only slightly impacts the
overall frequency of planets \citep{2007AA...468..721B,bergfors13}.  However, there
are hints that binary companions with separations $\lesssim100$\,AU, for
which these dynamical effects are enhanced, may reduce the number of giant
planets formed \citep[see e.g.][]{2007AA...468..721B,2007AA...474..273E,
wang14}
and leave a characteristic imprint on their properties 
\citep{2007AA...462..345D}.  However, the available data on such systems
is currently still too limited to confirm these hypotheses.  In any case,
the discovery of circumstellar planets in binary systems with separations
as small as $\sim$20\,AU \citep{2003ApJ...599.1383H,2011AA...528A...8C}, 
whose existence presents a considerable challenge to current planet 
formation theories \citep{2011CeMDA.111...29T}, proves that planet
formation can succeed even in dynamically excited environments.

In general, planetary orbits in a binary system are stable on long time
scales if the overall system architecture is strongly hierarchical
\citep{1999AJ....117..621H}.  In the case of circumstellar (s-type) 
planet orbits, this means that the planet's semimajor axis $a$ must
be smaller than a critical upper limit
$a\sub{c}$, which in turn is a fraction of the binary's separation 
$a\sub{b}$ depending on the binary's mass ratio and eccentricity. 
In sufficiently close binary systems 
($a\sub{b}\lesssim 5$--10\,AU), a second regime of
stable planet orbits become viable: \emph{circumbinary} (p-type) orbits,
which enclose the orbits of both stellar components of the binary host
system.  In this regime, the stability criterion takes the form of a
lower limit on the planet's semimajor axis, $a>a\sub{c}$, 
with $a\sub{c}/a\sub{b}$ ranging from 2 to 4.

Since the detectability of a planet with transit photometry and Doppler 
spectroscopy is highly biased towards small semimajor axes, circumbinary
planets are generally more challenging to detect with these techniques
than circumstellar planets.  Despite this disadvantage, seven transiting 
circumbinary planets have been discovered by the Kepler mission
\citep[e.g.,][]{2011Sci...333.1602D,2012Sci...337.1511O}, including a 
multi-planet system and a planet in the habitable zone, hinting at a 
rich and varied, yet largely undiscovered population of circumbinary planets
\citep{2013arXiv1308.6328W, armstrong14}.  Several other 
circumbinary planet candidates were discovered on the basis of eclipse
timing variations in post-common-envelope binary systems \citep[e.g.,][]
{2009AJ....137.3181L,2010AA...521L..60B}. Some of these claims
have been disputed \citep[e.g.,][]{2013MNRAS.435.2033H}, but for at 
least one target, NN~Ser, the evidence for circumbinary planets appears 
robust \citep{marsh14}. It has been 
suggested that at least some of these planets might have formed as
second-generation planets from ejected stellar material after the
common-envelope phase \citep{2010arXiv1001.0581P}.  Efforts are also 
made to search for circumbinary planets with Doppler spectroscopy
\citep{2009ApJ...704..513K}.

All of these indirect planet detection techniques are heavily
biased towards short-period planets, and do not provide meaningful 
constraints beyond orbital radii of 5--10\,AU.  On the other hand,
direct imaging powered by adaptive optics and differential imaging 
techniques has proven to be a powerful technique to probe the outer
reaches of stellar systems for giant planets 
\citep[e.g.][]{neuhaeuser03,2007ApJ...670.1367L,2012AA...544A...9V,2013AA...553A..60R,
2013ApJ...777..160B,2013ApJ...773...73J}, leading to a number of 
discoveries of planets
\citep[e.g.][]{2010Sci...329...57L,2010Natur.468.1080M,
2013ApJ...763L..32C,2013ApJ...774...11K,2013arXiv1310.7483R} and 
brown-dwarf companions \citep[e.g.][]{2009ApJ...707L.123T,
2010ApJ...720L..82B}.  However, the published imaging surveys
have largely avoided close binary targets, leaving a vast unexplored
territory in the parameter space of circumbinary planets.  
While \citet{crepp10} target narrow binaries in search of circumbinary
tertiary bodies, their sensitivity is limited to stellar companions.
\citet{2013A&A...553L...5D} reported the imaging discovery of a 
12--14\,$M\sub{Jup}$ substellar companion orbiting a pair of M-dwarf
stars, though the comparatively high mass ratio implies a stellar 
rather than planetary origin. To date, no \emph{bona fide} 
circumbinary planet has been imaged.

The SPOTS project (Search for Planets Orbiting Two Stars; 
\citealt{2013prpl.conf2K012T}) aims to fill in this gap by
conducting the first direct imaging survey dedicated to circumbinary
planets.  In a first stage, 26 nearby young spectroscopic binary 
targets have been observed over the past 2 years with VLT NaCo, and
41 more targets are foreseen for a second stage using VLT SPHERE in
the coming 2--3 years.  In this work, we present the scientific
justification, observing strategy, and preliminary results for the
first stage of the SPOTS survey.  Due to the unavailability of NaCo
in ESO periods P92 and P93, the follow-up efforts for this stage are
still ongoing; thus, final observational results and statistical 
analysis will be presented in a future publication.  Furthermore,
we complement our survey with a comprehensive census of the
body of archival high-contrast imaging data of spectroscopic 
binary targets \citepalias[\citealt{bonavita14}, herafter][]{bonavita14},
including a detailed statistical analysis based
on the \texttt{MESS} code \citep{2011arXiv1110.4917B}.

In Section~\ref{s:sciencecase}, we discuss the scientific merit of 
direct imaging of circumbinary planets in more detail.  
Section~\ref{s:survey} describes the survey design and lists the 
targets covered by the first stage of SPOTS.  We report on our
high-contrast observations and their preliminary results in 
Section~\ref{s:obs}, and summarize our conclusions in 
Section~\ref{s:conclusion}.


\section{The science case for direct imaging of circumbinary planets}
\label{s:sciencecase}

As outlined in \citet{2013prpl.conf2K012T}, the scientific motivation
of the SPOTS project to search for circumbinary planets by direct 
imaging can be summarized in the following four points.

\subsection{Unexplored territory}

Since stellar binarity is wide-spread, with approximately half of all
Sun-like stars having a stellar companion \citep{2010ApJS..190....1R}, 
circumbinary planets may
constitute a significant fraction of the overall planet population.
In the spirit of fundamental research, it is therefore inherently 
interesting to explore this largely unknown parameter space.  

The existence of circumbinary planets is favored by both observational
and theoretical arguments.  For instance, the existence of circumbinary
disks \citep{1992IAUS..151...21M, 1994AA...286..149D} and the growth and
crystalization of dust in such disks \citep{2008ApJ...673..477P, 
2004AA...427..651D} are well attested, and thus provide the necessary
ingredients for planet formation.  While secular excitation of planetesimal
orbits driven by the binary has been proposed as an obstacle to core 
accretion, favoring erosion over accretion \citep{2012ApJ...761L...7M}, 
\citet{2013ApJ...764L..16R} proposes that the asymmetric gas disk in such a 
system dampens the excitation and promotes accretion.  
\citet{2013A&A...553A..71M} instead finds that the overall effect of the
eccentric disk on the planetesimal population is destructive, whereas
\citet{2013ApJ...773...74M} find that a `dead zone' within a layered 
circumbinary disk may promote accretion, and even locally render planet 
formation more efficient than at similar separations around single stars.
In any case, the discovery of seven transiting circumbinary planets with
Kepler \citep{2013arXiv1308.6328W} demonstrates that planet formation does
successfully occur in close circumbinary configurations.  Furthermore, direct
imaging is sensitive to planets in wider orbits, where the perturbing 
influence of the central binary decreases.

\subsection{Good observability with direct imaging}
\label{s:goodobs}

Unlike Doppler spectroscopy and transit photometry, direct imaging does
not suffer technical difficulties from spectroscopic binary targets.  As
long as the binary is not resolved, it behaves like a single star for
the purposes of adaptive optics guidance and differential imaging 
methods.  Resolved binaries may achieve comparable performance if the
secondary star is significantly fainter than the primary, as observations
from our SPOTS pilot survey demonstrate (Section~\ref{s:obs}).  These
observational constraints coincide with the ideal conditions for 
stable circumbinary planet orbits (close stellar separation and/or high
stellar mass ratio; \citealt{1999AJ....117..621H}), and are therefore
compatible with most astrophysically promising targets.

Furthermore, circumbinary planets may offer a distinct advantage in terms of
contrast.  The number of massive planets formed in a system is thought to
scale with the available mass of the protoplanetary disk, which in turn
scales with the mass of the host star \citep[e.g][]{2008ApJ...673..502K}.
This notion is supported empirically by the fact that the majority of 
directly imaged planets have been found around early-type host
stars \citep{2010Sci...329...57L,2010Natur.468.1080M,2013ApJ...763L..32C,
2013arXiv1310.7483R}.
If this relation also holds for close binary stars, then a pair of G-type
stars should offer the same planet-forming resources as a single A-type.
However, the former system is several magnitudes fainter than the latter, 
and thus offers much more favorable conditions for direct imaging of 
self-luminous planets.

\subsection{More planets on wide orbits?}

Close binary systems offer two mechanisms that may increase the number
of planets on wide orbits, which are the favored targets of direct
imaging.

\begin{description}
\item[\emph{Scattering.}]
Planets forming or migrating close to the inner edge of the stability 
region may experience a close encounter with the host system's secondary
star, which results in dynamical scattering of the planet
\citep{2003MNRAS.345..233N,2004MNRAS.347..613V}.  While most
of these events end with the ejection of the planet from the system, 
some produce highly eccentric, bound orbits with large semimajor axes.
These planets may subsequently be damped into less eccentric
wide orbits through interaction with the protoplanetary disk.

\smallskip
Numerical simulations by \citet{pierens08} suggest that massive 
giant planets 
($\ge 1\,M_\mathrm{Jup}$), which make favorable objects for direct
imaging, are particularly prone to scattering, whereas lower-mass 
planets may instead remain stranded near the inner edge of the
stability region \citep[see also][]{pierens13,kley14}.  
This prediction has received some
observational support in a recent study by \citet{armstrong14} on 
Kepler transit photometry data, which indicates that the frequency 
of small planets (radius smaller than Jupiter's) on short orbits 
(periods $\le 300$\,d) is consistent between tight binary and single
host stars, but that Jovian-sized planets appear significantly 
depleted in the circumbinary case.  Considering that the high system
mass of binary systems is thought to favor the formation of massive
planets (cf.\ Section~\ref{s:goodobs}), these findings suggest a
selective depletion mechanism for such planets consistent with
\citet{pierens08}.

\item[\emph{Migration.}]
A second possibility is continuous outward migration as opposed to
isolated scattering events \citep{2007MNRAS.378.1589M}.  This requires
transfer of angular momentum from the host binary to a planet via 
viscous disk interaction, rendering the binary orbit tighter and the
planet orbit wider.  Due to the low mass ratio between the planet
and the stars and the non-linearity of the gravitational potential,
a small change in the binary orbit has a large effect on the planet
orbit, which may extend out to 20--50\,AU.
\end{description}

While the chances of either process occurring in a system may be low,
one must keep in mind that giant planets in wide orbits are rare to
begin with \citep{2010ApJ...717..878N}; thus these mechanisms may
still result in a measurable enrichment of the population
of such planets.

\subsection{Improved constraints on planet formation theories}

Circumbinary systems provide an excellent laboratory for testing theories
of planet formation and evolution.  Theoretical studies have predicted
both positive and negative effects of binarity on planet formation, in
both the core accretion scenario
\citep[e.g.,][]{2013A&A...553A..71M,2013ApJ...773...74M} and the
gravitational instability scenario \citep{2005MNRAS.363..641M,
2006ApJ...641.1148B}.  Such effects are expected to leave
characteristic imprints
on the planet demography around binary systems.  Measuring those
differences to the planet demography around single stars may then 
provide insights into the underlying physical processes.  Similar 
studies for circumstellar planets in wide binaries have already found
tentative evidence for such differences \citep[e.g.,][]
{2007AA...468..721B}.  

For circumbinary planets, a first indication of
such a demographic feature is found in Kepler transit
photometry data, where circumbinary planets have so far been found 
exclusively in host systems with binary orbital periods 
$P_\mathrm{bin} \le 7$\,d, although the detection bias from sampling
coverage favors target binaries with shorter periods 
\citep{2013arXiv1308.6328W, armstrong14}. Possible explanations
for this feature include reduced detectability of planets orbiting 
short-period binaries due to a larger spread of orbital misalignment, 
reduced planet formation efficiency due to higher secular 
perturbations in such systems, or loss of planets to dynamic scattering
in the formation phase of the short-period binary.
Sampling the circumbinary planet demography with direct imaging could
test whether this effect extends to wide planet orbits, and thus help
identify underlying astrophysical causes.


\section{Survey design}
\label{s:survey}


\subsection{Target selection}
\label{s:targets}

\begin{table*}[ptb]
\caption{Target list for the NaCo-based stage of the SPOTS survey.}
\label{tab:master}
\tiny
\centering
\begin{sideways}
\begin{tabular}{llr|rrrlr|rrrrr|l}
\\
\hline\hline 
 Name            &  RA(2000)   &  Dec(2000)   & Dist  &  Age   &    $H$    &   SpT  & $M_\mathrm{star}$     &  $M_\mathrm{comp}$  &  $\rho$    &  $P$     & $e$  & $a_\mathrm{crit}$  & Notes   \\
                 &             &              &  (pc) & (Myrs) &  (mag)  &        & ($M_{\odot}$)  & ($M_{\odot}$) &  (arcsecs) & (days) &      &  (AU)     &      \\   
\hline
HIP 9892 = HD 13183                  & 02:07 18.10 &  -53:11:56.00   &  50.9       &   30       &        &           &           &             &            &         &       &           &  \\
HIP 12545                  & 02:41:25.90 &  +05:59:18.40   &  42.0       &   16       & 7.20   & M...      & 0.76     &              &   --       &   --    &   --  &   $<20$   &  \\
UX~For = HIP 12716                  & 02 43 25.57 &  -37:55 42.53   &  40.7       &  600       & 6.10   &  G6V+     & 0.88     &  0.64       &            &   0.95     &   0.00  &   0.05    & Triple \\
                           &             &                 &              &            &        &           & 0.88+0.64 &  0.58       &   0.38     &   --    &   --  &   71.3    & Triple \\
HIP 16853 = HD 22705                 & 03:36:53.40 &  -49:57:28.90   &  41.7       &   30       & 6.26   &  G2V      &  1.0      &  0.40       &   --       &  201.0  & 0.00  &   1.77    &  \\ 
V1136~Tau = CHR~14 = HD 284163                  & 04 11 56.22 &  +23:38:10.76   &  38.1       &  625       &        &  K0       &0.78+0.49  &  0.52       &   0.31     &  40y    & 0.853 &  51.1     & Triple \\  
TYC 5907-1244-1            & 04:52:49.50 &  -19:55 02.00   &  72.0       &   30       & 7.50   &  K1Ve     &  0.87     &             &   --       &   --     &   --  &   $<10$  &   \\
AF~Lep = HIP 25486                  & 05:27:04.76 &  -11:54;03.47   &  27.0       &   16       & 2.93   &  F7V      &  1.06    & $0.76^a$    &   --       &   --     &   --  &   $<10$  &   \\
HIP 25709 = HD 36329         & 05:29:24.10 &  -34:30:56.00   &  70.9       &   30       &        &  G3V      &  1.07    &             &            &          &       &    $<10$  &   \\
XZ~Pic = HIP 27134                  & 05:45:13.40 &  -59:55:26.00   &  49.8       &  300       &        &  K0V(e)   &  0.95    &   --        &   --       &   --     &   --  &   $<1$   &   \\
Alhena = $\gamma$~Gem = HIP 31681 & 06:37:42.71 &  +16:23:57.41   &  33.5       &  300       &  1.84  &  A1.5IV+  & 3.0      &     1.0        &  0.273    &  12.63yr & 0.89  &     40     &  \\
TYC 8104 0991 1            & 06:39:05.70 &  -45:12:58.00   &  56.5       &  100       & 8.25   &  K3Ve     & 0.83      &    --         &    --        &     --     &   --       &          &  \\
26~Gem = HIP 32104                 & 06:42:24.33 &  +17:38:43.11   &  43.6       &   30       & 5.07   &  A2V      & 2.70      & 0.51       &   --       &    522   &  0.10 &  4.76      &  \\ 
EM Cha = RECX7                 & 08:43:07.24 &  -79:04:52.50   &  97.0   &    8       & 7.75   & K7Ve      & 1.0       & 0.4        &   --       &    2.6   &  --   &  0.10  &    \\
TYC 8569 3597 1            & 08:51:56.40 &  -53:55:57.00   & 141.0      &   30       &        &  G9V      &  1.11         &            &   --       &   24.06  &  --   &  $<1$  &    \\
GS~Leo = HIP 46637                  & 09:30:35.83 &   10:36:06.25   &  42.3        &  300       & 6.69   &   G5      & 0.89      &   --         &   --       &    --    &  --   &  $<1$  &  Wide comp. \\
TYC 9399 2452 1            & 09:19 24.00 &  -77:38:45.00   &  56.8        &  800       & 7.56   &  K0IVe    & 0.96          &   --          &  --   &          &  --      &  $<1$   &  Wide comp. \\
HIP 47760 = HD 84323                 & 09:44:14.10 &  -11:52:21.00   &  76.3       &   30       &        & G3V       &  1.08            &  --        &  --  &   --      &  --      &   $<10$   & \\ 
TYC 6604 0118 1            & 09:59:08.40 &  -22:39:35.00   &   61.1       &  100           & 7.49   & K2Ve      &   0.83   &    0.74  &  --  &   1.84    & 0.00         &  0.08         & \\
HS~Lup = HIP 74049                  & 15:07 57.75 &  -45:34:45.84   &  45.1       &  250       &        & G5IV+G5IV &   1.03           &            &  --  &  17.83    &  0.20    &           & \\
HIP 76629 = HD 133822     & 15:38:57.50 &  -57:42:27.00   &  38.5       &   16       &        & K0V       &    1.12          & $\sim$0.11         &  --    & $\sim$4.5 yr & $\sim$0.5    & $\sim$10.9   & Wide comp. \\
HIP 78416 = HD 143215     & 16:00 31.32 &  -36:05 16.62   &  84.39       &   16       &        & G1V       &   1.22           &            &   -- &   --      &  --      &  $<10$    & Wide comp. \\
BS~Ind = HIP 105404                & 21:20:59.80 &  -52:28:40.10   &  45.15       &   30       & 6.70   & K0V       & 0.90 & 0.80   &   -- &  1223     & 0.60     &  10.23    & Triple \\
IK Peg                     & 21:26:26.66 &  +19:22:32.30   &  46.36       &  100       &        & A8m+DA    &  1.45        &  0.99      &  --  &  21.72    & 0.00     &     0.45      & \\
$\delta$ Cap               & 21:47:02.44 &  -16:07:38.23   &  11.87       &  540       & 2.01   & A5        &  1.50        & 0.56       &  --  &  1.0     &  0.01     &  0.06     & \\
CS~Gru = HIP 109901                & 22:15:35.20 &  -39:00 51.00   &  56.10       &  100       & 7.12   & K0V       &  0.89        & 0.45       &  --  &   --     &  --       &  $<10$    &  \\
TYC 6386 0896 1            & 22:44:38.90 &  -15:56:29.00   &  57.9        &  150       & 7.47   & K1Ve      &  0.88            &            &  --  &   --     &  --       &  $<1$     &  \\
\end{tabular} 								        
\end{sideways} 
\end{table*}

The selection of targets was based on an extensive compilation of nearby 
young stars assembled in preparation of the SPHERE guaranteed-time survey. 
The indication of young age is derived from various indicators such as
membership to young moving groups, lithium content, chromospheric and coronal
emission, and fast rotation. 
From this list we selected 67 young spectroscopic and close visual 
binaries as suitable targets for a direct-imaging search for circumbinary 
planets. As a selection criterion, we demanded that the background-limited
minimum detectable planet mass in a one-hour NaCo observation be 
$\le 5 M_\mathrm{Jup}$. This imposes joint constraints on the distance and
age of the target systems, allowing higher ages for more nearby stars.  For
early-type target stars, which are expected to form higher-mass planets than 
Sun-like stars, this criterion was relaxed to $\le 8 M_\mathrm{Jup}$.  The
median minimum detectable planet mass was below $1 M_\mathrm{Jup}$ for the
Moving Groups subsample, $2.8 M_\mathrm{Jup}$ for the Field Stars 
subsample, and $4.4 M_\mathrm{Jup}$ for the Early-Type subsample.
A total of 26 targets were eventually observed in the NaCo stage of the
SPOTS survey.

It should be noted that close tidally-locked binaries are characterized by
high levels of chromospheric and coronal emission and fast rotation even at
old ages, mimicking the appearence of young stars. Indeed, several old 
tidally-locked binaries are included in previous direct-imaging surveys
\citepalias{bonavita14}. To exclude this kind of systems, which
are not suited for our science aim, we required that the estimate of young
age not be based solely on activity and rotation, but also on other indicators
such as lithium content and group membership, unless the availability of the 
binary orbital elements (e.g., long period) allowed to rule out the
tidal-locking scenario. We should mention, however, that lithium content 
can also be altered in close binaries, making the age estimate challenging
\citep{1992A&A...253..185P}.

Table \ref{tab:master} summarizes the stellar and binary parameters of the
observed target sample.
The determination of the stellar parameters follows the methods described in
\citet{2014arXiv1405.1559D}. We listed the mass of the binary companion and the orbital
elements where available. We also derived the critical semimajor axis for 
dynamical stability \citep{1999AJ....117..621H}, to have an estimate of the 
minimum separation at which circumbinary planets can be found in stable orbits.
In most cases, the results were significantly smaller than the inner working
angle of our imaging observations.
Where orbital elements are not available, upper limits based on, e.g., the 
observed radial velocity variability are listed.
The targets are also described individually in Appendix~\ref{s:appendix}.


\subsection{Observing strategy}
\label{s:strategy}

For the first stage of the SPOTS survey, we make use of the Very Large
Telescope (VLT) NAOS-Conica (NaCo) high-contrast adaptive-optics
(AO) imaging facility.
We follow the 
observing technique that was used for the NaCo Large Program 
(\citealt{2014arXiv1405.1560C}; ESO programs 184.C-0567, 089.C-0137, 
090.C-0252), a direct imaging
survey for planets around single stars, whose observational
stage has been completed successfully and whose concluding publications
are currently in preparation.  Analogous observing strategies are 
employed by other surveys as well, such as the ongoing SEEDS survey on 
Subaru HiCIAO \citep{tamura09}.

For each survey target, an exploratory $H$-band (1.6\,$\mu$m) 
high-contrast imaging observation is taken on VLT NaCo.  
If the data reduction reveals faint point sources at significance
levels of $\ge 5\,\sigma$ with convincing visual characteristics in 
the vicinity of the target star, a second epoch of observation must be 
taken in order to test these companion candidates for common proper 
motion with the target star.  From our experience with the
NaCo Large Program, we expect roughly 30\% of all targets to have
candidates in need of such astrometric follow-up \citep{2014arXiv1405.1560C}.
The time between the exploratory and 
follow-up observations is chosen such that the star exhibits a 
measurable amount of proper and parallactic motion 
(ideally $\ge 100$\,mas) with respect to the sky background.  Due to
the slow duty cycle involving proposal-writing and service-mode
observation,
the practical minimum epoch difference is roughly one year, which is a
sufficient amount of time for most target stars.  
Should any candidates
be confirmed as co-moving and thus as physical companions to their
stars, further follow-up observations will be pursued.

The exploratory and astrometric follow-up observations follow the same
pattern.  Each observing block is one hour long, and includes target
acquisition, two short four-point dithering sequences of unsaturated
images taken with the \texttt{ND\_short} neutral density filter 
(transmission $0.0123\pm0.0005$ in $H$-band, \citealt{bonnefoy13})
for photometric calibration
purposes bracketing the science observations, and a sky background
observation for fixed-pattern noise calibration towards the end of the
observing block.  We use the $H$-band filter and the high-resolution
camera with a pixel scale of 13.7\,mas and a field of view of 1024 
pixels = 14\arcsec{} across.  For two particularly bright targets, 
the field of view was reduced by half in order to allow for faster
detector readout times.
The NAOS AO system is run in natural
guide star mode and must provide close to diffraction-limited resolution
in order to qualify an observing run as successful.  Pupil-tracking 
mode is used in order to allow for angular differential imaging 
(ADI; \citealt{marois06}) data reduction, and cube-mode data storage
is employed to reduce overhead.  Due to the limited Strehl ratio in 
$H$-band as well as unreliable pointing stability, we forego the use of
a coronagraph and instead saturate the target star out to $\sim7$
pixels in order to achieve high background-limited sensitivity.  The
integrated exposure time on the science target is on the order of 
40\% of the duration of the observing block.

All data sets of this kind are reduced by subtracting a fixed-pattern
noise, dividing by a flat field image, binning the ensemble of exposures into
a manageable number of master frames if necessary, and finally 
subtracting the stellar point-spread function (PSF) with the ADI
technique using the LOCI algorithm (Locally Optimized Combination of
Images; \citealt{lafreniere07}).  Since the aim is to detect 
point sources, the LOCI parameters are set to aggressive values
(tyically, a frame selection criterion of $N_\delta=0.5$\,FWHM and an
optimization area of $N_A = 300$ PSF footprints).
Fake planet signals are injected into
the data in order to measure and compensate for the loss of planet 
flux due to partial self-subtraction, following the procedure in
\citet{lafreniere07}.  We estimate the hypothetical
companion masses for the candidates by comparing their absolute 
$H$-band magnitude with the \texttt{COND} evolutionary models
\citep{allard01,2003AA...402..701B}.

Companion candidates are identified as faint signals with morphologies
compatible with the PSF core and statistical significance of at least
$5\,\sigma$.  The significance is calculated as the S/N ratio, using a 
radial noise profile calculated as the standard deviation in concentric
annuli around the star.  

While the NaCo-based first stage of the SPOTS survey is planned to be
concluded in 2015, a second stage covering the remaining 40 targets is
foreseen to be executed with VLT SPHERE in the coming years.


\section{Observations and preliminary results}
\label{s:obs}

\begin{figure*}[pt]
    \centering
    \includegraphics[width=\linewidth]{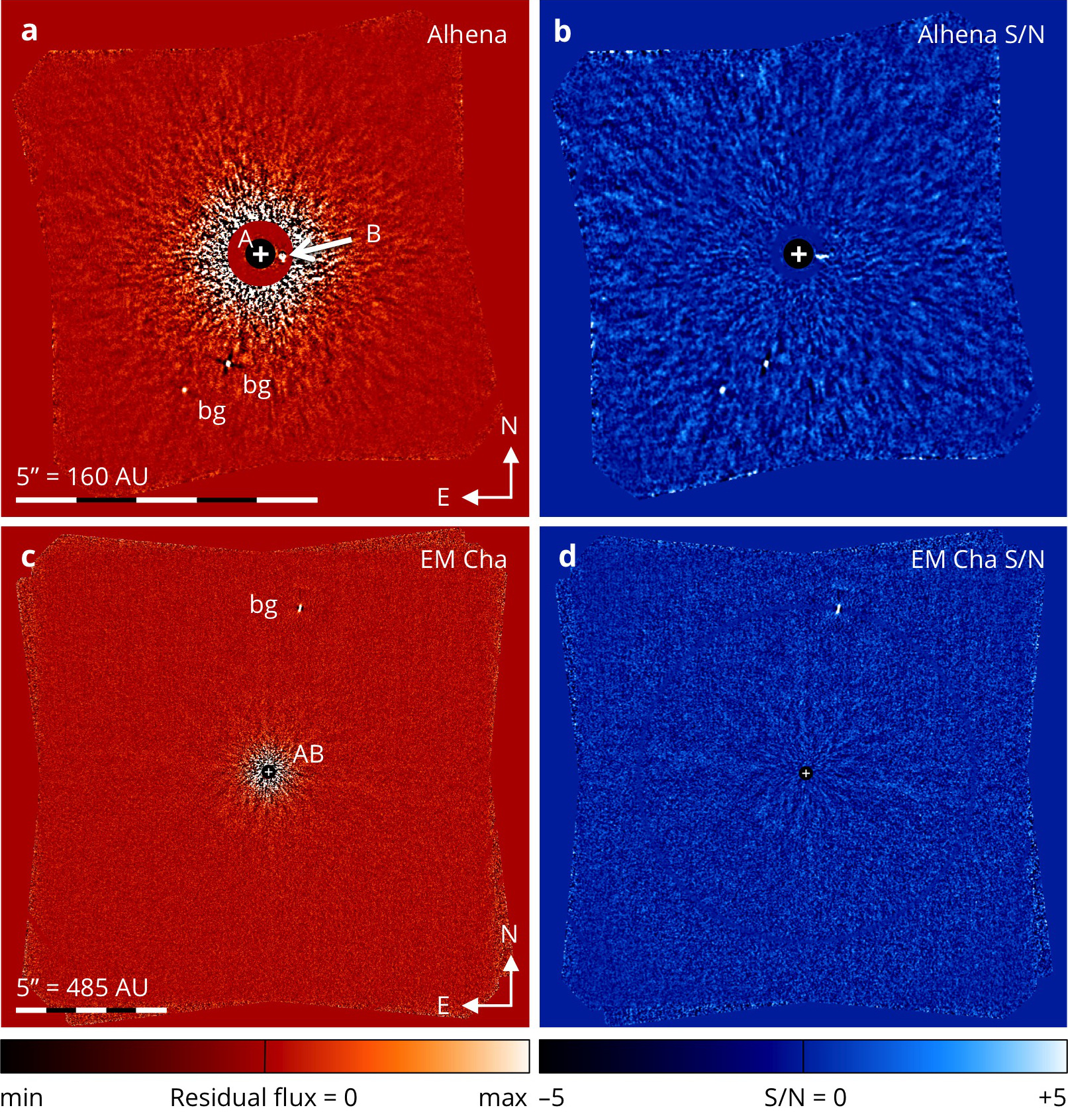}
    \caption{Sample high-contrast $H$-band images from the SPOTS 
        pilot survey
        taken on VLT NaCo. \textbf{(a)} LOCI ADI image of Alhena at
        linear stretch of $\pm7\times10^{-7}$ times the primary
        star's peak flux.  The image is centered on the A-type primary
        star (marked with a plus sign and the label \textsf{A}).  
        The area within 0\farcs55 of the
        primary is displayed with a software attenuation
        of a factor of $1/1000$ in order to render the resolved 
        G-type secondary star (marked \textsf{B}) visible.  Despite
        the presence of the secondary, the ADI processing effectively
        removes the primary's PSF, revealing two
        faint point sources nearby.  Our follow-up observations 
        identify these sources as background stars (marked \textsf{bg}).
        If they had been co-moving companions sharing the host
        system's age, their $H$-band brightness
        would have corresponded to masses of 29\,$M_\mathrm{Jup}$ 
        and 45\,$M_\mathrm{Jup}$ 
        in the \texttt{COND} evolutionary models by 
        \citet{2003AA...402..701B}.
        The field of view read out from the detector was reduced for 
        this observation to allow for faster readout times.
        \textbf{(b)} The signal-to-noise (S/N) ratio map corresponding
        to image (a), produced by cutting the image into concentric 
        annuli and dividing their pixel values by their standard deviation,
        displayed at a linear stretch of $[-5,5]\,\sigma$. Since the 
        secondary star is much fainter than the primary, it does not 
        disturb the well-behaved residual speckle noise pattern.
        \textbf{(c)} LOCI ADI image of EM~Cha at a linear stretch of 
        $\pm2.2\times10^{-5}$ times the unresolved binary's peak flux. 
        The location of the 
        unresolved target binary is marked by a plus sign and the label
        \textsf{AB}.  A known background star is visible 
        (marked \textsf{bg}).  Its brightness corresponds to a 
        7\,$M_\mathrm{Jup}$ planet at the system age.
        \textbf{(d)}  The S/N map corresponding to image (c), 
        demonstrating well-behaved residual noise.
        }
    \vspace*{2cm}
    \label{f:images}
\end{figure*}

\begin{table*}[tbp]
\centering
\begin{tabular}{l l r c l c c}
  &     & Integr.  & Bin. & Candidates \& bkgd.\ obj.\
     & \multicolumn{2}{c}{Follow-up}\\
Target & Epoch & time (s) & res.? & (hypothetical masses) & needed? & obtained?\\  
\hline
\multicolumn{6}{l}{\textit{ESO Program 088.C-0291(A)}\phantom{\LARGE !} \smallskip}\\
$\delta$ Cap & 2011-10-06 & 1320 & & --- &  & \\
IK Peg & 2011-10-10 & 1938 & & 2 $\times$ 7\,$M_\mathrm{Jup}$ planet at large separation & $\blacksquare$ & $\square$\\
    & & & & 1 $\times$ 7\,$M_\mathrm{Jup}$ planet near IWA (4.9\,$\sigma$)\\
HIP 12545 & 2011-10-31 & 2040 & & --- & & \\
TYC 5907 1244 1 & 2011-11-05 & 1224 & & --- & & \\
HIP 16853 & 2011-11-09 & 1596 & & 1 $\times$ star & & \\
HIP 9892 & 2011-12-03 & 960 & & 1 $\times$ background galaxy & & \\
EM Cha & 2011-12-15 & 1428 & & 1 $\times$ confirmed background star & & \\
TYC 8569 3597 1 & 2011-12-17 & 1200 & & crowded field & $\blacksquare$ & \\
AF~Lep & 2011-12-20 & 1596 & & --- & & \\
Alhena & 2011-12-22 & 1320 & $\blacksquare$ & 2 $\times$ brown dwarf & $\blacksquare$ & $\blacksquare$\\
HIP 25709 & 2011-12-22 & 2520 & & --- & & \\
26~Gem & 2012-01-13 & 1638 & &  1 $\times$ high-mass brown dwarf at IWA & 
    $\blacksquare$ & $\blacksquare$\\
HIP 76629 & 2012-07-20 & 1672 & & crowded field & $\blacksquare$ & \\
\\
\multicolumn{6}{l}{\textit{ESO program 090.C-0416(B)}
    \smallskip}\\
V1136 Tau & 2012-11-21 & 1560 & $\blacksquare$ & 1 $\times$ star & & \\
TYC 6386 0896 1 & 2012-11-26 & 1224 & & --- & & \\
XZ~Pic & 2013-01-03 & 960 & & --- & & \\
UX~For & 2013-01-05 & 2280 & $\blacksquare$ & --- & & \\
TYC 8104 0991 1 & 2013-01-07 & 1200 & & --- & & \\
HIP 47760 & 2013-01-24 & 1224 & & --- & & \\
TYC 6604 0118 1 & 2013-01-26 & 1224 & & --- & & \\
GS~Leo & 2013-02-10 & 1560 & & 1 $\times$ 12\,$M_\mathrm{Jup}$ planet near IWA & $\blacksquare$ & $\blacksquare$\\
TYC 9399 2452 1 & 2013-03-03 & 1224 & & 1 $\times$ 10\,$M_\mathrm{Jup}$ planet & $\blacksquare$ & \\
HS~Lup & 2013-04-30 & 1596 & & 3 $\times$ 6--11\,$M_\mathrm{Jup}$ planet & $\blacksquare$ & \\
HIP 78416 & 2013-05-11 & 720 & & 1 $\times$ foreground star & $\blacksquare$ & \\
    & & & & 4 $\times$ 5--11\,$M_\mathrm{Jup}$ planet & & \\
    & & & & 3 $\times$ background galaxy & & \\
CS~Gru & 2013-06-26 & 720 & & --- \\
BS~Ind & 2013-06-28 & 1560 & & 2 $\times$ 3--9\,$M_\mathrm{Jup}$ planet
    & $\blacksquare$ & \\
\end{tabular}
\bigskip
\caption{Status of SPOTS exploratory observations. A black square in the
column ``Bin.\ res.?''\ indicates that the target binary has been resolved
in our images.  In the last two columns,
a black square indicates that follow-up observations were deemed necessary,
and were successfully executed, respectively.  The empty square for the 
target IK~Peg marks a follow-up observations that was executed, but
turned out to be insufficiently sensitive to re-detect the candidate
sources.  The acronym IWA stands for inner working angle, which, in this
particular context, refers to the minimum angular separation from the star
at which there is sufficient field rotation to perform LOCI ADI processing.
Note that detections near the IWA suffer from increased false alarm 
probability due to small-number statistics \citep{mawet14}.
Background 
galaxies are identified by their visibly extended morphology that is
inconsistent with the stellar PSF observed in the unsaturated calibration
frames. None of
the targets with reliable follow-up observations have yielded any 
co-moving companions.}
\label{t:exobs}
\end{table*}

\subsection{Exploratory observations}

\begin{figure}[pt]
    \centering
    \includegraphics[width=\linewidth,trim=10mm 0mm 0mm 0mm]{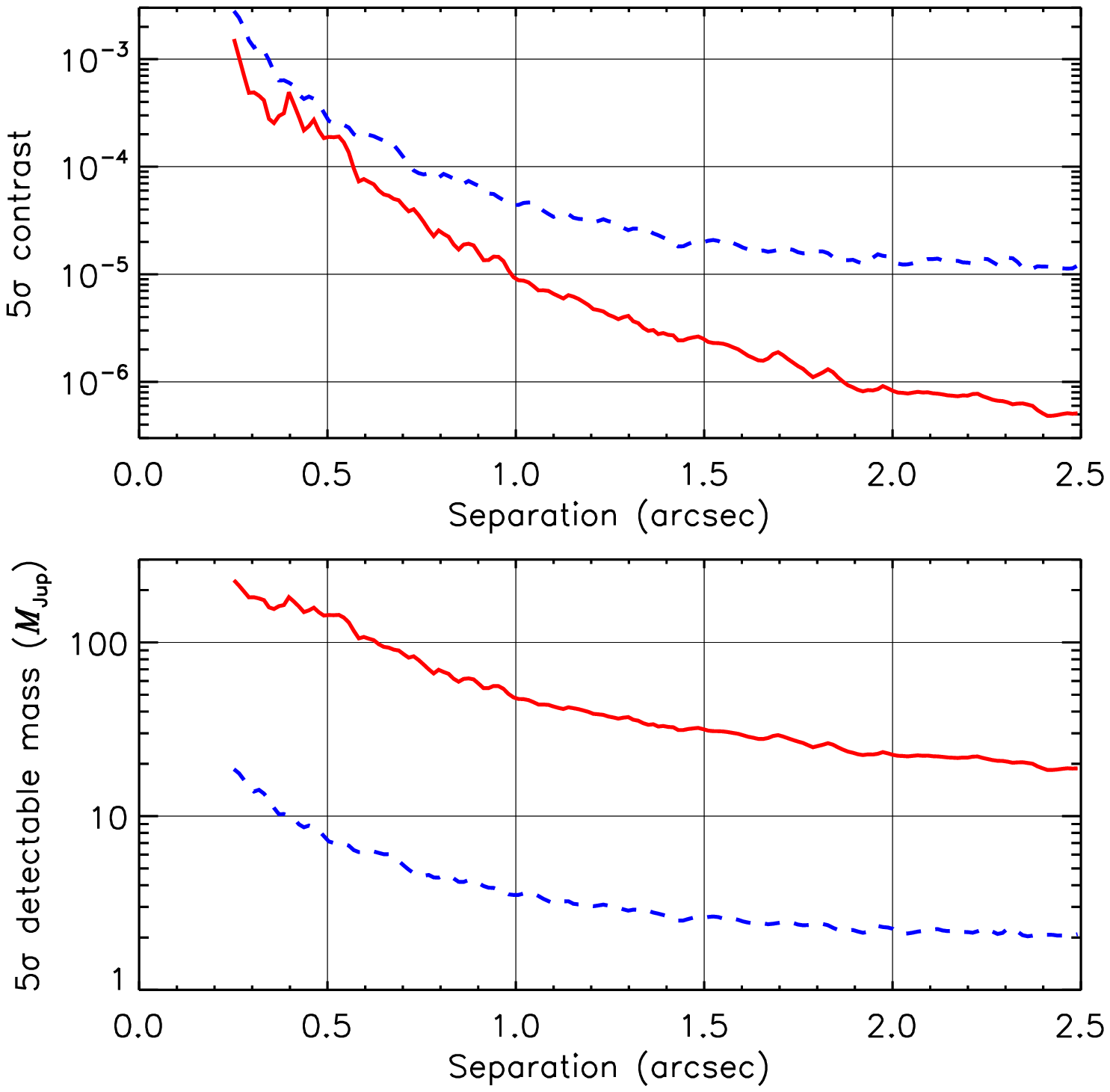}
    \caption{Plot of the 5\,$\sigma$ contrast (top panel) and the
        corresponding 5\,$\sigma$ minimum detectable planet mass (bottom panel)
        for the LOCI ADI images from the two sample targets Alhena 
        (red solid lines) and EM~Cha
        (blue dashed lines).  Both curves are corrected for the partial
        loss of planet flux due to self-subtraction in the data reduction.
        The total amount of field rotation captured was low in both 
        observations (11.8$^\circ$ for Alhena and 12.6$^\circ$ for EM~Cha);
        thus, the curves are conservative representations of the survey's
        sensitivity. 
        Note that Alhena provides higher contrast than EM Cha since it
        is a brighter star and therefore offers more dynamic range between
        the stellar peak flux and the background sensitivity limit.  On
        the other hand, since EM~Cha is a much younger system than Alhena
        (8 Myr vs.\ 300 Myr), the minimum detectable companion masses are 
        lower for EM Cha by an order of magnitude.
        }
    \label{f:contrast}
\end{figure}

A total of 14 and 13 targets were observed with VLT NaCo 
as part of the SPOTS survey under the ESO programs 
088.C-0291(A) and 090.C-0416(B), 
respectively.  Details for these observations are presented in 
Table~\ref{t:exobs}.
For each observed target, the epoch of first observation,
the integrated time on target, and the presence of candidates or other
notable field objects is noted.  The last two columns furthermore note
whether the exploratory observations revealed candidates in need of 
follow-up observations, and whether such follow-up has been obtained
successfully at the time of writing.  Figure~\ref{f:images} 
illustrates the high quality of the high-contrast data on the examples
of the two targets Alhena and EM~Cha, and Figure~\ref{f:contrast}
provides plots of the contrast and minimum detectable planet mass 
curves for those two targets.

\subsection{Follow-up observations}

In ESO program 090.C-0416(A), four targets from the P88 exploratory 
observations flagged for follow-up were observed for a second epoch.
In the case of IK~Peg, the 
background-limited sensitivity of the follow-up data was 
insufficient to re-detect the faint candidates of the first epoch, 
probably due to worse weather conditions during the second epoch.
This observation will therefore have to be repeated.


For the P90 target GS~Leo, we pursued special follow-up observations
via director's discretionary time (DDT) on the Subaru IRCS facility
\citep{IRCS}.  This target lies in the Northern hemisphere and thus 
passes close to zenith when viewed from the Mauna Kea observatory, 
affording a much higher field rotation rate and thus superior ADI
performance and inner working angle (IWA)\footnote{%
    While the term IWA originates from coronography, we use it to 
    refer to the minimum angular separation from the star at which
    the total field rotation is sufficient to perform LOCI ADI for
    a given dataset.
} than from Paranal. In the
first-epoch data, the candidate appeared as a 6\,$\sigma$ signal
at a separation of 0\farcs4 from the star, very close to the inner 
working angle given by the poor field rotation range of those data.
Although the follow-up data provided twice the contrast performance
of the first-epoch data at the location of the candidate and offered
a significantly smaller inner working angle, the candidate signal was
not re-detected.  We conclude that the candidate was a deviant residual
speckle from the ADI data reduction in the first epoch, rather than a
real on-sky source.  At the IWA, the number of frames available to the
ADI algorithm for PSF subtraction is
extremely limited, thus the residual noise can no longer be assumed to
approximate Gaussian statistics as is commonly the case in the 
well-sampled parts of an ADI output image \citep{mawet14}.  The 
apparent 6\,$\sigma$ significance in the first-epoch image may 
therefore have been overestimated.

\subsection{Preliminary results from the planet search}
\label{s:prelimres}

Since many targets whose exploratory observations revealed promising 
companion candidates remain without a second epoch and therefore 
cannot yet be tested for common proper motion, we defer an in-depth
report and statistical analysis of the results to a future paper. 
Since NaCo is scheduled to be offered again on VLT in ESO period P94,
the remaining follow-up observations could be concluded by early 2015
at the earliest.

In the meantime, the current body of observations constitute a 
successful pilot run for the full SPOTS survey, demonstrating the 
feasibility and validity of our observational strategy.  Most of our
targets are unresolved spectroscopic binaries, which behave like
single stars for the purpose of observation and data reduction.  As
the images of Alhena in Figure~\ref{f:images} demonstrate, even targets
with a resolved secondary star are useful to search for planets with
ADI, in particular if the secondary is much fainter than the primary.
Of our 26 exploratory targets, 10 have been found to require follow-up,
which coincides well with the follow-up rate of 30\% found
for single stars in the NaCo large program \citep{2014arXiv1405.1560C}. 


Among the three targets with reliable follow-up observations (Alhena,
26~Gem, GS~Leo), no co-moving companion has been found.  A total
of seven targets with at least one candidate for a planetary-mass companion
remain to be confirmed (IK Peg, TYC 8569 3597 1, HIP 76629, TYC 9399 2452 1,
HS~Lup, HIP 78416, BS~Ind).

The
final astrometric analyses will be published in the full survey paper
foreseen for 2015.

\subsection{Astrometry of resolved binary targets}
\label{s:binastro}

\begin{figure}[pt]
    \centering
    \includegraphics[width=0.95\linewidth]{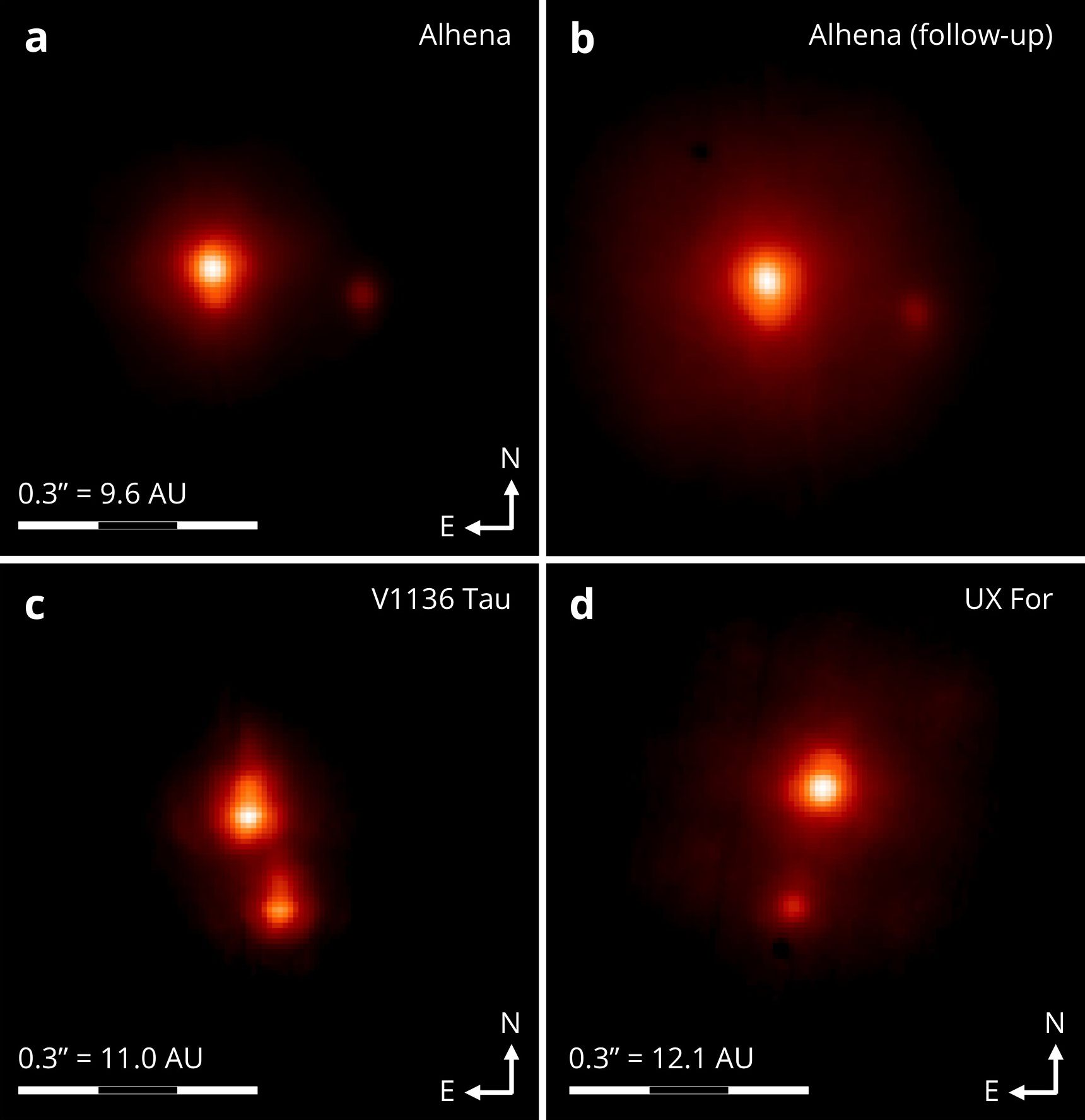}
    \caption{Resolved multiple targets from the NaCo-based stage of the SPOTS
        survey: \textbf{(a)} Alhena ($= \gamma$ Gem; exploratory observation),
        \textbf{(b)}
        Alhena (follow-up observation), \textbf{(c)} V1136~Tau, suffering some
        degraded AO correction,
        \textbf{(d)} UX~For, showing the bright unresolved AB binary and
        the fainter C tertiary. All images are shown at a logarithmic 
        stretch spanning 2.2 orders of magnitude.
        }
    \label{f:resbin}
\end{figure}

\begin{figure*}[bpt]
    \centering
    \includegraphics[height=4.8cm]{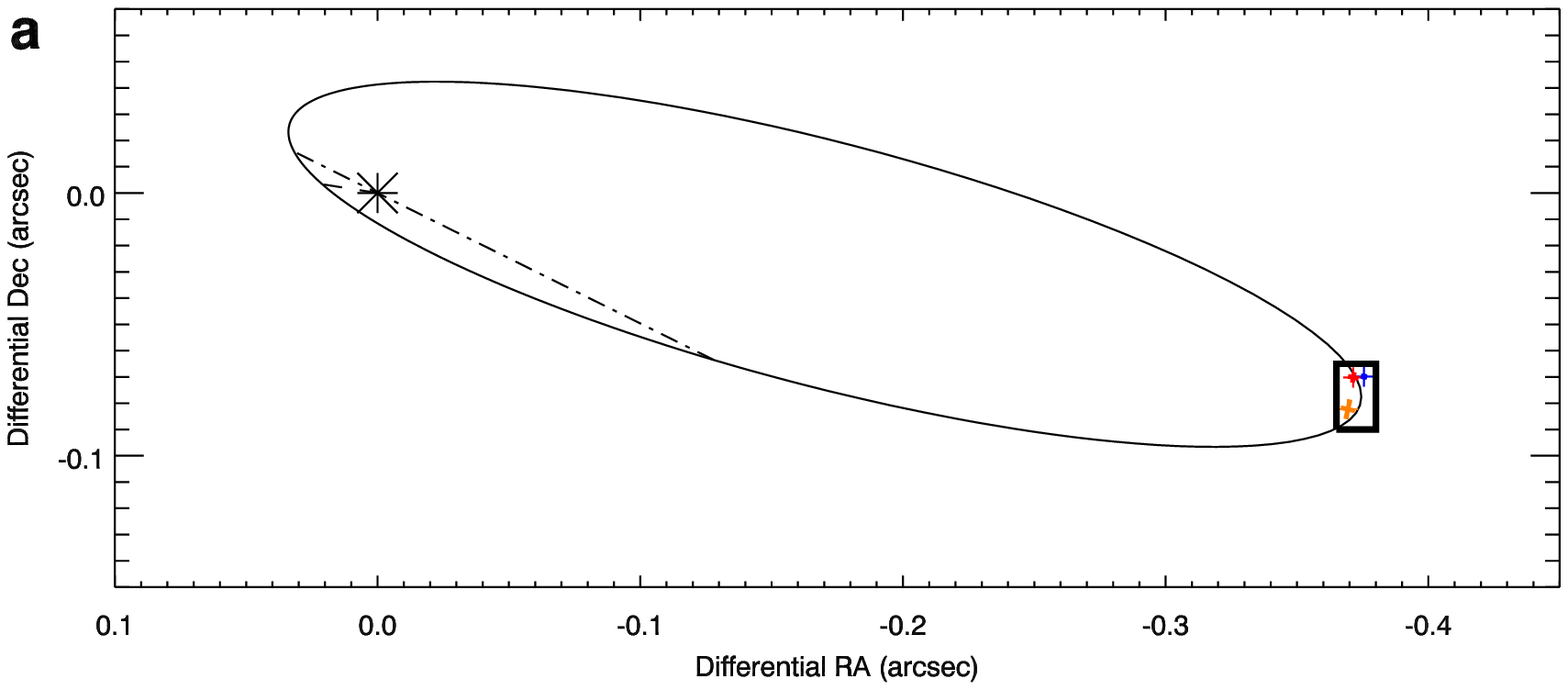}
    \includegraphics[height=4.8cm]{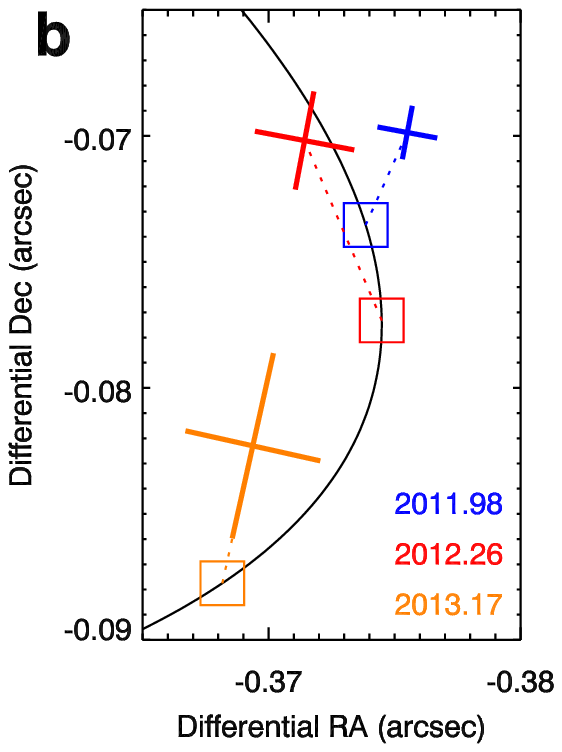}
    \includegraphics[height=4.8cm]{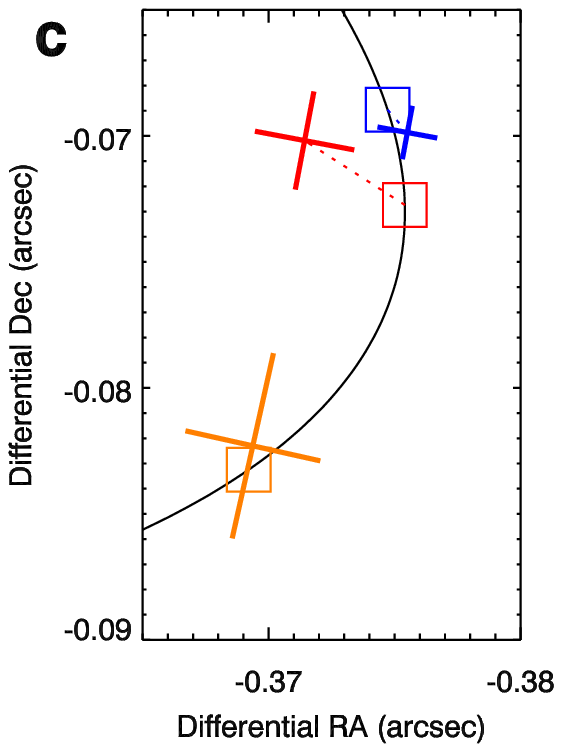}
    \caption{Astrometry of the resolved binary target Alhena
        ($= \gamma$ Gem). \textbf{(a)} Overview the orbit of 
        Alhena B relative to Alhena A, 
        as reported in \citet{drummond14}. The black rectangular
        frame indicates the area shown at larger magnification in the
        following panels. The position of Alhena A is marked with a
        star.
        \textbf{(b)} Observed relative positions of
        Alhena B (crosses) and their predicted positions (squares) on
        the assumed orbit for each of the three epochs (2011.98, 
        2013.17: this work; 2012.26: \citealt{drummond14}).  The sizes
        of the crosses correspond to the astrometric error bars in the
        azimuthal and radial directions. Note that the observed 
        positions are systematically and significantly offset to the 
        North ($\chi^2=31.4$). \textbf{(c)} The same for an adjusted
        orbit with $\Omega^\prime=\Omega + 0\fdg7$, leading to a
        good fit the data ($\chi^2=6.8$).}
    \label{f:alhena_astro}
\end{figure*}


As an auxiliary science result, our survey data provide accurate 
stellar astrometry for three resolved target binary systems: Alhena
(two epochs), V1136~Tau, and UX~For.  Unsaturated images from these
four observations are shown in Figure~\ref{f:resbin}, demonstrating the
clear detections.  Each of these targets is briefly discussed in a 
dedicated paragraph below. The numerical results are summarized in 
Table~\ref{t:binastro}.

For all targets, unsaturated frames taken as part of our high-contrast
observing runs for the purpose of photometric calibration were used to
measure the relative astrometry of the target binary.  The position of
each stellar component was extracted using the \texttt{gcntrd} function
in \texttt{IDL}.  The error of this measurement was estimated to the
FWHM of the secondary's PSF divided by its S/N ratio.  A second error
contribution is due to the uncertainty in True North orientation and
plate scale.  Since the SPOTS observations are performed in service
mode and distributed over the course of years, it was impractical to
include regular astrometric calibration observations.  For the 
first-epoch observation of Alhena, we use the calibration values 
reported for a roughly contemporary epoch in the NaCo Large Program
(\citealt{2014arXiv1405.1560C}; Table 5). For all epochs after early 2012,
we adopt the last reported values (from 2012-01-02), which carry a 
considerable uncertainty in True North orientation.  We represent this
uncertainty with an error bar of 0\farcs4, spanning the full range of 
True North orientations reported between 2009-11-23 and 2012-01-02.
The final errors on the position angle and separation of the 
secondary star are obtained by combining the centroiding and 
calibration errors in quadrature.


\begin{table}[]
\centering
\begin{tabular}{@{}llr@{~$\pm$~}lr@{~$\pm$~}l@{}}
Target          & Epoch         & \multicolumn{2}{l@{}}{Sep.\ (mas)}       
    & \multicolumn{2}{l@{}}{PA ($^\circ$)}\\
\hline
Alhena\phantom{\Large!}        & 2011-12-22    & 381.9 & 1.2   & 259.46 & 0.16\\
                & 2013-03-02   & 378.4 & 2.7    & 257.44 & 0.57\\
V1136~Tau          & 2012-11-22   & 247.8 & 0.5    & 197.74 & 0.40\\
UX~For       & 2013-01-02   & 305.8 & 0.9    & 165.32 & 0.43
\vspace*{2mm}
\end{tabular}
\caption{Relative astrometry of the three resolved target binaries in
the NaCo-based stage of the SPOTS survey.}
\label{t:binastro}
\end{table}

\medskip
\noindent \textbf{Alhena} ($= \gamma$ Gem, WDS J06377+1624Aa,Ab,
HD 47105, HIP 31681) comprises an A-type primary
and a G-type secondary.  While constraints from Doppler spectroscopy 
and astrometric tracking of the system's photocenter had been available
for decades, the secondary star was only recently resolved for the 
first time \citep{drummond14}, allowing these authors to propose a first
complete orbital solution.  Our exploratory and follow-up observations
add two new epochs of relative astrometry, bracketing the single epoch
of that author.  In Figure~\ref{f:alhena_astro}, we illustrate the
relative orbit of Alhena~B around A as proposed in \citet{drummond14},
and the three resolved astrometric measurements. All data points appear to be
systematically and significantly offset from their expected locations
to the North ($\chi^2=31.4$). While refitting the orbit to all 
published constraints is beyond the scope of this work, we note that 
adding $+0\fdg7$ to the longitude of the ascending node, $\Omega$, 
removes these discrepancies and yields a good fit ($\chi^2=6.8$)
without impacting the Doppler spectroscopy constraints 
(Figure~\ref{f:alhena_astro}c).


\medskip
\noindent \textbf{V1136~Tau} ($=$ CHR~14, HD~284163, WDS J04119+2338AB, HIP 19591) 
is listed with a period of $P=40.9$\,yr, a semi-major axis of 
$a=0\farcs31$, and an eccentricity of $e=0.853$ in \citet{malkov12}.
Both component stars appear elongated towards the North in our images,
likely due to an imperfect PSF shape caused by wave-front sensing on 
the resolved, low-contrast binary guide star.  Our astrometry places
the secondary star between the positions reported for a bracketing 
pair of epochs by \citet{tokovinin12, tokovinin14}.

\medskip
\noindent \textbf{UX~For} ($=$ HIP~12716, HD~17084, TOK 187) is a 
triple system comprising an unresolved G6V-type spectroscopic binary 
with a 0.955-day period and a resolved tertiary component first 
reported in \citet{hartkopf12}. No orbital solution is given, though
the period of the tertiary is estimated to 40 years.


\section{Conclusion}
\label{s:conclusion}

SPOTS (Search for Planets Orbiting Two Stars) is the first dedicated 
high-contrast imaging survey aimed at sampling the population of 
giant planets on wide circumbinary orbits.  In this work (Paper I), 
we have presented our scientific rationale for this kind of survey, 
which can be summarized in four points:
\begin{itemize}
\item Theoretical and observational work predicts that giant planets
can form, evolve, and survive on long time scales in circumbinary 
environments.  However, most direct imaging surveys so far have 
excluded binaries from their target lists, leaving this interesting
demographic of planets largely unexplored. 
\item Close binary systems that make promising hosts for circumbinary
planets do not cause difficulties in direct imaging observations or
data reduction.  Furthermore, a binary host star may offer more 
planet-forming resources than a single star of the same overall
brightness.
\item Circumbinary planets may be scattered or continuously pushed out
to larger orbital radii by dynamic interaction with their host binary,
enriching the population of wide-orbit planets.
\item Since binarity of the host star influences planet formation by
core accretion and gravitational instability in different ways, 
sampling the circumbinary planet demography may yield new and unique
insights into the processes of planet formation.
\end{itemize}

Furthermore, we have presented the design and current status of the
SPOTS survey.  In the context of our pilot survey on VLT NaCo, we have
obtained exploratory observations of 26 targets, and are currently in
the process of concluding the follow-up efforts for the promising 
companion candidates in that sample.  Our preliminary results 
demonstrate the feasibility and validity of our observing strategy.
No confirmed co-moving companion has been discovered so far; however,
seven targets with promising candidates remain to be re-observed
for common proper motion testing (cf.\ Section~\ref{s:prelimres}).

Finally, we have reported on the astrometry of the three resolved 
binaries in our observed sample (Alhena, V1136~Tau, UX~For).

In a separate publication \citepalias{bonavita14}, 
we present a statistical analysis of the
direct imaging constraints on tight binary targets collected from the
target lists of several published surveys.

A full report and statistical analysis for this NaCo-based stage of
the SPOTS survey will be delivered in a third publication upon 
completion of the follow-up observations (expected in 2015).  A 
second stage of the SPOTS survey comprising 40 more targets is then
foreseen with VLT SPHERE in the following years.


\begin{acknowledgements}
The authors thank David Lafreni\`ere for providing the source 
code for his LOCI algorithm, and the anonymous referee for useful 
comments.
CTh is supported by the European Commission under the
Marie Curie IEF grant No.~329875, and JCa by the U.S.\ National Science
Foundation under award No.~1009203.
SD and MB ackowledge support from PRIN-INAF "Planetary systems at young 
ages". The authors thank the ESO Paranal staff for their assistance and
execution of service-mode observations.
The authors wish to recognize and acknowledge the very significant 
cultural role and reverence that the summit of Mauna Kea has always 
had within the indigenous Hawaiian community.  We are most fortunate 
to have the opportunity to conduct observations from this mountain.
\end{acknowledgements}



\bibliographystyle{aa}
\bibliography{cb}

\begin{thebibliography}{113}
\expandafter\ifx\csname natexlab\endcsname\relax\def\natexlab#1{#1}\fi

\bibitem[{{Allard} {et~al.}(2001){Allard}, {Hauschildt}, {Alexander},
  {Tamanai}, \& {Schweitzer}}]{allard01}
{Allard}, F., {Hauschildt}, P.~H., {Alexander}, D.~R., {Tamanai}, A., \&
  {Schweitzer}, A. 2001, \apj, 556, 357

\bibitem[{{Armstrong} {et~al.}(2014){Armstrong}, {Osborn}, {Brown}, {Faedi},
  {G\'omez Maqueo Chew}, {Martin}, {Pollacco}, \& {Udry}}]{armstrong14}
{Armstrong}, D.~J., {Osborn}, H., {Brown}, D., {et~al.} 2014, ArXiv e-prints

\bibitem[{{Artymowicz} \& {Lubow}(1994)}]{artymowicz94}
{Artymowicz}, P. \& {Lubow}, S.~H. 1994, \apj, 421, 651

\bibitem[{{Bailey} {et~al.}(2012){Bailey}, {White}, {Blake}, {Charbonneau},
  {Barman}, {Tanner}, \& {Torres}}]{2012ApJ...749...16B}
{Bailey}, III, J.~I., {White}, R.~J., {Blake}, C.~H., {et~al.} 2012, \apj, 749,
  16

\bibitem[{{Baraffe} {et~al.}(2003){Baraffe}, {Chabrier}, {Barman}, {Allard}, \&
  {Hauschildt}}]{2003AA...402..701B}
{Baraffe}, I., {Chabrier}, G., {Barman}, T.~S., {Allard}, F., \& {Hauschildt},
  P.~H. 2003, A\&A, 402, 701

\bibitem[{{Bender} \& {Simon}(2008)}]{2008ApJ...689..416B}
{Bender}, C.~F. \& {Simon}, M. 2008, \apj, 689, 416

\bibitem[{{Bergfors} {et~al.}(2013){Bergfors}, {Brandner}, {Daemgen}, {Biller},
  {Hippler}, {Janson}, {Kudryavtseva}, {Gei{\ss}ler}, {Henning}, \&
  {K{\"o}hler}}]{bergfors13}
{Bergfors}, C., {Brandner}, W., {Daemgen}, S., {et~al.} 2013, \mnras, 428, 182

\bibitem[{{Beuermann} {et~al.}(2010){Beuermann}, {Hessman}, {Dreizler},
  {Marsh}, {Parsons}, {Winget}, {Miller}, {Schreiber}, {Kley}, {Dhillon},
  {Littlefair}, {Copperwheat}, \& {Hermes}}]{2010AA...521L..60B}
{Beuermann}, K., {Hessman}, F.~V., {Dreizler}, S., {et~al.} 2010, A\&A, 521,
  L60

\bibitem[{{Biller} {et~al.}(2010){Biller}, {Liu}, {Wahhaj}, {Nielsen}, {Close},
  {Dupuy}, {Hayward}, {Burrows}, {Chun}, {Ftaclas}, {Clarke}, {Hartung},
  {Males}, {Reid}, {Shkolnik}, {Skemer}, {Tecza}, {Thatte}, {Alencar},
  {Artymowicz}, {Boss}, {de Gouveia Dal Pino}, {Gregorio-Hetem}, {Ida},
  {Kuchner}, {Lin}, \& {Toomey}}]{2010ApJ...720L..82B}
{Biller}, B.~A., {Liu}, M.~C., {Wahhaj}, Z., {et~al.} 2010, \apjl, 720, L82

\bibitem[{{Biller} {et~al.}(2013){Biller}, {Liu}, {Wahhaj}, {Nielsen},
  {Hayward}, {Males}, {Skemer}, {Close}, {Chun}, {Ftaclas}, {Clarke}, {Thatte},
  {Shkolnik}, {Reid}, {Hartung}, {Boss}, {Lin}, {Alencar}, {de Gouveia Dal
  Pino}, {Gregorio-Hetem}, \& {Toomey}}]{2013ApJ...777..160B}
{Biller}, B.~A., {Liu}, M.~C., {Wahhaj}, Z., {et~al.} 2013, \apj, 777, 160

\bibitem[{{Bonavita} {et~al.}(2011){Bonavita}, {Chauvin}, {Desidera},
  {Gratton}, {Janson}, {Beuzit}, {Kasper}, \&
  {Mordasini}}]{2011arXiv1110.4917B}
{Bonavita}, M., {Chauvin}, G., {Desidera}, S., {et~al.} 2011, ArXiv e-prints

\bibitem[{{Bonavita} \& {Desidera}(2007)}]{2007AA...468..721B}
{Bonavita}, M. \& {Desidera}, S. 2007, A\&A, 468, 721

\bibitem[{{Bonavita} {et~al.}(2014){Bonavita}, {Desidera}, {Thalmann}, {Vigan},
  {Chauvin}, {Lafreni\`{e}re}, \& {Jayawardhana}}]{bonavita14}
{Bonavita}, M., {Desidera}, S., {Thalmann}, C., {et~al.} 2014, in prep. (Paper
  II)

\bibitem[{{Bonnefoy} {et~al.}(2013){Bonnefoy}, {Boccaletti}, {Lagrange},
  {Allard}, {Mordasini}, {Beust}, {Chauvin}, {Girard}, {Homeier}, {Apai},
  {Lacour}, \& {Rouan}}]{bonnefoy13}
{Bonnefoy}, M., {Boccaletti}, A., {Lagrange}, A.-M., {et~al.} 2013, \aap, 555,
  A107

\bibitem[{{Boss}(2006)}]{2006ApJ...641.1148B}
{Boss}, A.~P. 2006, \apj, 641, 1148

\bibitem[{{Carson} {et~al.}(2013){Carson}, {Thalmann}, {Janson}, {Kozakis},
  {Bonnefoy}, {Biller}, {Schlieder}, {Currie}, {McElwain}, {Goto}, {Henning},
  {Brandner}, {Feldt}, {Kandori}, {Kuzuhara}, {Stevens}, {Wong}, {Gainey},
  {Fukagawa}, {Kuwada}, {Brandt}, {Kwon}, {Abe}, {Egner}, {Grady}, {Guyon},
  {Hashimoto}, {Hayano}, {Hayashi}, {Hayashi}, {Hodapp}, {Ishii}, {Iye},
  {Knapp}, {Kudo}, {Kusakabe}, {Matsuo}, {Miyama}, {Morino}, {Moro-Martin},
  {Nishimura}, {Pyo}, {Serabyn}, {Suto}, {Suzuki}, {Takami}, {Takato},
  {Terada}, {Tomono}, {Turner}, {Watanabe}, {Wisniewski}, {Yamada}, {Takami},
  {Usuda}, \& {Tamura}}]{2013ApJ...763L..32C}
{Carson}, J., {Thalmann}, C., {Janson}, M., {et~al.} 2013, \apjl, 763, L32

\bibitem[{{Chauvin} {et~al.}(2011){Chauvin}, {Beust}, {Lagrange}, \&
  {Eggenberger}}]{2011AA...528A...8C}
{Chauvin}, G., {Beust}, H., {Lagrange}, A.-M., \& {Eggenberger}, A. 2011, A\&A,
  528, A8

\bibitem[{{Chauvin} {et~al.}(2014){Chauvin}, {Vigan}, {Bonnefoy}, {Desidera},
  {Bonavita}, {Mesa}, {Boccaletti}, {Buenzli}, {Carson}, {Delorme},
  {Hagelberg}, {Montagnier}, {Mordasini}, {Quanz}, {Segransan}, {Thalmann},
  {Beuzit}, {Biller}, {Covino}, {Feldt}, {Girard}, {Gratton}, {Henning},
  {Kasper}, {Lagrange}, {Messina}, {Meyer}, {Mouillet}, {Moutou}, {Reggianni},
  {Schlieder}, \& {Zurlo}}]{2014arXiv1405.1560C}
{Chauvin}, G., {Vigan}, A., {Bonnefoy}, M., {et~al.} 2014, ArXiv e-prints

\bibitem[{{Christian} {et~al.}(2002){Christian}, {Mathioudakis}, \&
  {Vennes}}]{2002IBVS.5281....1C}
{Christian}, D.~J., {Mathioudakis}, M., \& {Vennes}, S. 2002, Information
  Bulletin on Variable Stars, 5281, 1

\bibitem[{{Covino} {et~al.}(1997){Covino}, {Alcala}, {Allain}, {Bouvier},
  {Terranegra}, \& {Krautter}}]{1997A&A...328..187C}
{Covino}, E., {Alcala}, J.~M., {Allain}, S., {et~al.} 1997, \aap, 328, 187

\bibitem[{{Crepp} {et~al.}(2010){Crepp}, {Serabyn}, {Carson}, {Ge}, \&
  {Kravchenko}}]{crepp10}
{Crepp}, J., {Serabyn}, E., {Carson}, J., {Ge}, J., \& {Kravchenko}, I. 2010,
  \apj, 715, 1533

\bibitem[{{Cutispoto} {et~al.}(2002){Cutispoto}, {Pastori}, {Pasquini}, {de
  Medeiros}, {Tagliaferri}, \& {Andersen}}]{2002A&A...384..491C}
{Cutispoto}, G., {Pastori}, L., {Pasquini}, L., {et~al.} 2002, \aap, 384, 491

\bibitem[{{Cutispoto} {et~al.}(1999){Cutispoto}, {Pastori}, {Tagliaferri},
  {Messina}, \& {Pallavicini}}]{1999A&AS..138...87C}
{Cutispoto}, G., {Pastori}, L., {Tagliaferri}, G., {Messina}, S., \&
  {Pallavicini}, R. 1999, \aaps, 138, 87

\bibitem[{{Dall} {et~al.}(2007){Dall}, {Foellmi}, {Pritchard}, {Lo Curto},
  {Allende Prieto}, {Bruntt}, {Amado}, {Arentoft}, {Baes}, {Depagne},
  {Fernandez}, {Ivanov}, {Koesterke}, {Monaco}, {O'Brien}, {Sarro}, {Saviane},
  {Scharw{\"a}chter}, {Schmidtobreick}, {Sch{\"u}tz}, {Seifahrt}, {Selman},
  {Stefanon}, \& {Sterzik}}]{2007A&A...470.1201D}
{Dall}, T.~H., {Foellmi}, C., {Pritchard}, J., {et~al.} 2007, \aap, 470, 1201

\bibitem[{{Davis} {et~al.}(2010){Davis}, {Kolb}, \&
  {Willems}}]{2010MNRAS.403..179D}
{Davis}, P.~J., {Kolb}, U., \& {Willems}, B. 2010, \mnras, 403, 179

\bibitem[{{Delorme} {et~al.}(2013){Delorme}, {Gagn{\'e}}, {Girard}, {Lagrange},
  {Chauvin}, {Naud}, {Lafreni{\`e}re}, {Doyon}, {Riedel}, {Bonnefoy}, \&
  {Malo}}]{2013A&A...553L...5D}
{Delorme}, P., {Gagn{\'e}}, J., {Girard}, J.~H., {et~al.} 2013, \aap, 553, L5

\bibitem[{{Desidera} \& {Barbieri}(2007)}]{2007AA...462..345D}
{Desidera}, S. \& {Barbieri}, M. 2007, A\&A, 462, 345

\bibitem[{{Desidera} {et~al.}(2014){Desidera}, {Covino}, {Messina}, {Carson},
  {Hagelberg}, {Schlieder}, {Biazzo}, {Alcala}, {Chauvin}, {Vigan}, {Beuzit},
  {Bonavita}, {Bonnefoy}, {Delorme}, {D'Orazi}, {Esposito}, {Feldt}, {Girardi},
  {Gratton}, {Henning}, {Lagrange}, {Lanzafame}, {Launhardt}, {Marmier},
  {Melo}, {Meyer}, {Mouillet}, {Moutou}, {Segransan}, {Udry}, \&
  {Zaidi}}]{2014arXiv1405.1559D}
{Desidera}, S., {Covino}, E., {Messina}, S., {et~al.} 2014, ArXiv e-prints

\bibitem[{{Desidera} {et~al.}(2006){Desidera}, {Gratton}, {Lucatello},
  {Claudi}, \& {Dall}}]{2006A&A...454..553D}
{Desidera}, S., {Gratton}, R.~G., {Lucatello}, S., {Claudi}, R.~U., \& {Dall},
  T.~H. 2006, \aap, 454, 553

\bibitem[{{Doyle} {et~al.}(2011){Doyle}, {Carter}, {Fabrycky}, {Slawson},
  {Howell}, {Winn}, {Orosz}, {Pr\v{s}a}, {Welsh}, {Quinn}, {Latham}, {Torres},
  {Buchhave}, {Marcy}, {Fortney}, {Shporer}, {Ford}, {Lissauer}, {Ragozzine},
  {Rucker}, {Batalha}, {Jenkins}, {Borucki}, {Koch}, {Middour}, {Hall},
  {McCauliff}, {Fanelli}, {Quintana}, {Holman}, {Caldwell}, {Still},
  {Stefanik}, {Brown}, {Esquerdo}, {Tang}, {Furesz}, {Geary}, {Berlind},
  {Calkins}, {Short}, {Steffen}, {Sasselov}, {Dunham}, {Cochran}, {Boss},
  {Haas}, {Buzasi}, \& {Fischer}}]{2011Sci...333.1602D}
{Doyle}, L.~R., {Carter}, J.~A., {Fabrycky}, D.~C., {et~al.} 2011, Science,
  333, 1602

\bibitem[{{Drummond}(2014)}]{drummond14}
{Drummond}, J.~D. 2014, \aj, 147, 65

\bibitem[{{Duch{\^e}ne} {et~al.}(2004){Duch{\^e}ne}, {Bouvier}, {Bontemps},
  {Andr{\'e}}, \& {Motte}}]{2004AA...427..651D}
{Duch{\^e}ne}, G., {Bouvier}, J., {Bontemps}, S., {Andr{\'e}}, P., \& {Motte},
  F. 2004, A\&A, 427, 651

\bibitem[{{Dutrey} {et~al.}(1994){Dutrey}, {Guilloteau}, \&
  {Simon}}]{1994AA...286..149D}
{Dutrey}, A., {Guilloteau}, S., \& {Simon}, M. 1994, A\&A, 286, 149

\bibitem[{{Eggenberger} {et~al.}(2007){Eggenberger}, {Udry}, {Chauvin},
  {Beuzit}, {Lagrange}, {S{\'e}gransan}, \& {Mayor}}]{2007AA...474..273E}
{Eggenberger}, A., {Udry}, S., {Chauvin}, G., {et~al.} 2007, A\&A, 474, 273

\bibitem[{{Evans}(1961)}]{1961RGOB...30...93E}
{Evans}, D.~S. 1961, Royal Greenwich Observatory Bulletins, 30, 93

\bibitem[{{Gagn{\'e}} {et~al.}(2014){Gagn{\'e}}, {Lafreni{\`e}re}, {Doyon},
  {Malo}, \& {Artigau}}]{2014ApJ...783..121G}
{Gagn{\'e}}, J., {Lafreni{\`e}re}, D., {Doyon}, R., {Malo}, L., \& {Artigau},
  {\'E}. 2014, \apj, 783, 121

\bibitem[{{Galland} {et~al.}(2005){Galland}, {Lagrange}, {Udry}, {Chelli},
  {Pepe}, {Queloz}, {Beuzit}, \& {Mayor}}]{2005AA...443..337G}
{Galland}, F., {Lagrange}, A.-M., {Udry}, S., {et~al.} 2005, A\&A, 443, 337

\bibitem[{{Griffin} \& {Gunn}(1981)}]{1981AJ.....86..588G}
{Griffin}, R.~F. \& {Gunn}, J.~E. 1981, \aj, 86, 588

\bibitem[{{Guenther} {et~al.}(2005){Guenther}, {Covino}, {Alcal{\'a}},
  {Esposito}, \& {Mundt}}]{2005AA...433..629G}
{Guenther}, E.~W., {Covino}, E., {Alcal{\'a}}, J.~M., {Esposito}, M., \&
  {Mundt}, R. 2005, A\&A, 433, 629

\bibitem[{{Guenther} \& {Esposito}(2007)}]{2007astro.ph..1293G}
{Guenther}, E.~W. \& {Esposito}, E. 2007, ArXiv Astrophysics e-prints

\bibitem[{{Haghighipour}(2010)}]{2010ASSL..366.....H}
{Haghighipour}, N., ed. 2010, Astrophysics and Space Science Library, Vol. 366,
  {Planets in Binary Star Systems}

\bibitem[{{Hartkopf} {et~al.}(2012{\natexlab{a}}){Hartkopf}, {Tokovinin}, \&
  {Mason}}]{hartkopf12}
{Hartkopf}, W.~I., {Tokovinin}, A., \& {Mason}, B.~D. 2012{\natexlab{a}}, \aj,
  143, 42

\bibitem[{{Hartkopf} {et~al.}(2012{\natexlab{b}}){Hartkopf}, {Tokovinin}, \&
  {Mason}}]{2012AJ....143...42H}
{Hartkopf}, W.~I., {Tokovinin}, A., \& {Mason}, B.~D. 2012{\natexlab{b}}, \aj,
  143, 42

\bibitem[{{Hatzes} {et~al.}(2003){Hatzes}, {Cochran}, {Endl}, {McArthur},
  {Paulson}, {Walker}, {Campbell}, \& {Yang}}]{2003ApJ...599.1383H}
{Hatzes}, A.~P., {Cochran}, W.~D., {Endl}, M., {et~al.} 2003, \apj, 599, 1383

\bibitem[{{Helt} \& {Jensen}(1989)}]{1989IBVS.3306....1H}
{Helt}, B.~E. \& {Jensen}, K.~S. 1989, Information Bulletin on Variable Stars,
  3306, 1

\bibitem[{{Holman} \& {Wiegert}(1999)}]{1999AJ....117..621H}
{Holman}, M.~J. \& {Wiegert}, P.~A. 1999, AJ, 117, 621

\bibitem[{{Horner} {et~al.}(2013){Horner}, {Wittenmyer}, {Hinse}, {Marshall},
  {Mustill}, \& {Tinney}}]{2013MNRAS.435.2033H}
{Horner}, J., {Wittenmyer}, R.~A., {Hinse}, T.~C., {et~al.} 2013, \mnras, 435,
  2033

\bibitem[{{Janson} {et~al.}(2013){Janson}, {Brandt}, {Moro-Mart{\'{\i}}n},
  {Usuda}, {Thalmann}, {Carson}, {Goto}, {Currie}, {McElwain}, {Itoh},
  {Fukagawa}, {Crepp}, {Kuzuhara}, {Hashimoto}, {Kudo}, {Kusakabe}, {Abe},
  {Brandner}, {Egner}, {Feldt}, {Grady}, {Guyon}, {Hayano}, {Hayashi},
  {Hayashi}, {Henning}, {Hodapp}, {Ishii}, {Iye}, {Kandori}, {Knapp}, {Kwon},
  {Matsuo}, {Miyama}, {Morino}, {Nishimura}, {Pyo}, {Serabyn}, {Suenaga},
  {Suto}, {Suzuki}, {Takahashi}, {Takami}, {Takato}, {Terada}, {Tomono},
  {Turner}, {Watanabe}, {Wisniewski}, {Yamada}, {Takami}, \&
  {Tamura}}]{2013ApJ...773...73J}
{Janson}, M., {Brandt}, T.~D., {Moro-Mart{\'{\i}}n}, A., {et~al.} 2013, \apj,
  773, 73

\bibitem[{{Jones} {et~al.}(2002){Jones}, {Paul Butler}, {Marcy}, {Tinney},
  {Penny}, {McCarthy}, \& {Carter}}]{jones02}
{Jones}, H.~R.~A., {Paul Butler}, R., {Marcy}, G.~W., {et~al.} 2002, \mnras,
  337, 1170

\bibitem[{{Kennedy} \& {Kenyon}(2008)}]{2008ApJ...673..502K}
{Kennedy}, G.~M. \& {Kenyon}, S.~J. 2008, \apj, 673, 502

\bibitem[{{Kley} \& {Haghighipour}(2014)}]{kley14}
{Kley}, W. \& {Haghighipour}, N. 2014, \aap, 564, A72

\bibitem[{{Kobayashi} {et~al.}(2000){Kobayashi}, {Tokunaga}, {Terada}, {Goto},
  {Weber}, {Potter}, {Onaka}, {Ching}, {Young}, {Fletcher}, {Neil},
  {Robertson}, {Cook}, {Imanishi}, \& {Warren}}]{IRCS}
{Kobayashi}, N., {Tokunaga}, A.~T., {Terada}, H., {et~al.} 2000, in Society of
  Photo-Optical Instrumentation Engineers (SPIE) Conference Series, Vol. 4008,
  Optical and IR Telescope Instrumentation and Detectors, ed. M.~{Iye} \& A.~F.
  {Moorwood}, 1056--1066

\bibitem[{{Konacki} {et~al.}(2009){Konacki}, {Muterspaugh}, {Kulkarni}, \&
  {He{\l}miniak}}]{2009ApJ...704..513K}
{Konacki}, M., {Muterspaugh}, M.~W., {Kulkarni}, S.~R., \& {He{\l}miniak},
  K.~G. 2009, \apj, 704, 513

\bibitem[{{Kuzuhara} {et~al.}(2013){Kuzuhara}, {Tamura}, {Kudo}, {Janson},
  {Kandori}, {Brandt}, {Thalmann}, {Spiegel}, {Biller}, {Carson}, {Hori},
  {Suzuki}, {Burrows}, {Henning}, {Turner}, {McElwain}, {Moro-Mart{\'{\i}}n},
  {Suenaga}, {Takahashi}, {Kwon}, {Lucas}, {Abe}, {Brandner}, {Egner}, {Feldt},
  {Fujiwara}, {Goto}, {Grady}, {Guyon}, {Hashimoto}, {Hayano}, {Hayashi},
  {Hayashi}, {Hodapp}, {Ishii}, {Iye}, {Knapp}, {Matsuo}, {Mayama}, {Miyama},
  {Morino}, {Nishikawa}, {Nishimura}, {Kotani}, {Kusakabe}, {Pyo}, {Serabyn},
  {Suto}, {Takami}, {Takato}, {Terada}, {Tomono}, {Watanabe}, {Wisniewski},
  {Yamada}, {Takami}, \& {Usuda}}]{2013ApJ...774...11K}
{Kuzuhara}, M., {Tamura}, M., {Kudo}, T., {et~al.} 2013, \apj, 774, 11

\bibitem[{{Lafreni{\`e}re} {et~al.}(2007{\natexlab{a}}){Lafreni{\`e}re},
  {Doyon}, {Marois}, {Nadeau}, {Oppenheimer}, {Roche}, {Rigaut}, {Graham},
  {Jayawardhana}, {Johnstone}, {Kalas}, {Macintosh}, \&
  {Racine}}]{2007ApJ...670.1367L}
{Lafreni{\`e}re}, D., {Doyon}, R., {Marois}, C., {et~al.} 2007{\natexlab{a}},
  ApJ, 670, 1367

\bibitem[{{Lafreni{\`e}re} {et~al.}(2007{\natexlab{b}}){Lafreni{\`e}re},
  {Marois}, {Doyon}, {Nadeau}, \& {Artigau}}]{lafreniere07}
{Lafreni{\`e}re}, D., {Marois}, C., {Doyon}, R., {Nadeau}, D., \& {Artigau},
  {\'E}. 2007{\natexlab{b}}, \apj, 660, 770

\bibitem[{{Lagrange} {et~al.}(2010){Lagrange}, {Bonnefoy}, {Chauvin}, {Apai},
  {Ehrenreich}, {Boccaletti}, {Gratadour}, {Rouan}, {Mouillet}, {Lacour}, \&
  {Kasper}}]{2010Sci...329...57L}
{Lagrange}, A.-M., {Bonnefoy}, M., {Chauvin}, G., {et~al.} 2010, Science, 329,
  57

\bibitem[{{Lee} {et~al.}(2009){Lee}, {Kim}, {Kim}, {Koch}, {Lee}, {Kim}, \&
  {Park}}]{2009AJ....137.3181L}
{Lee}, J.~W., {Kim}, S.-L., {Kim}, C.-H., {et~al.} 2009, ApJ, 137, 3181

\bibitem[{{Lyo} {et~al.}(2003){Lyo}, {Lawson}, {Mamajek}, {Feigelson}, {Sung},
  \& {Crause}}]{2003MNRAS.338..616L}
{Lyo}, A.-R., {Lawson}, W.~A., {Mamajek}, E.~E., {et~al.} 2003, \mnras, 338,
  616

\bibitem[{{Makarov}(2007)}]{2007ApJ...654L..81M}
{Makarov}, V.~V. 2007, \apjl, 654, L81

\bibitem[{{Makarov} \& {Kaplan}(2005)}]{2005AJ....129.2420M}
{Makarov}, V.~V. \& {Kaplan}, G.~H. 2005, ApJ, 129, 2420

\bibitem[{{Malkov} {et~al.}(2012{\natexlab{a}}){Malkov}, {Tamazian}, {Docobo},
  \& {Chulkov}}]{malkov12}
{Malkov}, O.~Y., {Tamazian}, V.~S., {Docobo}, J.~A., \& {Chulkov}, D.~A.
  2012{\natexlab{a}}, \aap, 546, A69

\bibitem[{{Malkov} {et~al.}(2012{\natexlab{b}}){Malkov}, {Tamazian}, {Docobo},
  \& {Chulkov}}]{2012A&A...546A..69M}
{Malkov}, O.~Y., {Tamazian}, V.~S., {Docobo}, J.~A., \& {Chulkov}, D.~A.
  2012{\natexlab{b}}, \aap, 546, A69

\bibitem[{{Malo} {et~al.}(2013){Malo}, {Doyon}, {Lafreni{\`e}re}, {Artigau},
  {Gagn{\'e}}, {Baron}, \& {Riedel}}]{2013ApJ...762...88M}
{Malo}, L., {Doyon}, R., {Lafreni{\`e}re}, D., {et~al.} 2013, \apj, 762, 88

\bibitem[{{Marcy} {et~al.}(2005){Marcy}, {Butler}, {Fischer}, {Vogt}, {Wright},
  {Tinney}, \& {Jones}}]{marcy05}
{Marcy}, G., {Butler}, R.~P., {Fischer}, D., {et~al.} 2005, Progress of
  Theoretical Physics Supplement, 158, 24

\bibitem[{{Marois} {et~al.}(2006){Marois}, {Lafreni{\`e}re}, {Doyon},
  {Macintosh}, \& {Nadeau}}]{marois06}
{Marois}, C., {Lafreni{\`e}re}, D., {Doyon}, R., {Macintosh}, B., \& {Nadeau},
  D. 2006, \apj, 641, 556

\bibitem[{{Marois} {et~al.}(2010){Marois}, {Zuckerman}, {Konopacky},
  {Macintosh}, \& {Barman}}]{2010Natur.468.1080M}
{Marois}, C., {Zuckerman}, B., {Konopacky}, Q.~M., {Macintosh}, B., \&
  {Barman}, T. 2010, \nat, 468, 1080

\bibitem[{{Marsh} {et~al.}(2014){Marsh}, {Parsons}, {Bours}, {Littlefair},
  {Copperwheat}, {Dhillon}, {Breedt}, {Caceres}, \& {Schreiber}}]{marsh14}
{Marsh}, T.~R., {Parsons}, S.~G., {Bours}, M.~C.~P., {et~al.} 2014, \mnras,
  437, 475

\bibitem[{{Martin} {et~al.}(2013){Martin}, {Armitage}, \&
  {Alexander}}]{2013ApJ...773...74M}
{Martin}, R.~G., {Armitage}, P.~J., \& {Alexander}, R.~D. 2013, \apj, 773, 74

\bibitem[{{Martin} {et~al.}(2007){Martin}, {Lubow}, {Pringle}, \&
  {Wyatt}}]{2007MNRAS.378.1589M}
{Martin}, R.~G., {Lubow}, S.~H., {Pringle}, J.~E., \& {Wyatt}, M.~C. 2007,
  \mnras, 378, 1589

\bibitem[{{Marzari} {et~al.}(2013){Marzari}, {Thebault}, {Scholl}, {Picogna},
  \& {Baruteau}}]{2013A&A...553A..71M}
{Marzari}, F., {Thebault}, P., {Scholl}, H., {Picogna}, G., \& {Baruteau}, C.
  2013, \aap, 553, A71

\bibitem[{{Mathieu}(1992)}]{1992IAUS..151...21M}
{Mathieu}, R.~D. 1992, in IAU Symposium, Vol. 151, Evolutionary Processes in
  Interacting Binary Stars, ed. {Y.~Kondo, R.~Sistero, \& R.~S.~Polidan}, 21

\bibitem[{{Mawet} {et~al.}(2014){Mawet}, {Milli}, {Wahhaj}, {Pelat}, {Absil},
  {Delacroix}, {Boccaletti}, {Kasper}, {Kenworthy}, {Marois}, {Mennesson}, \&
  {Pueyo}}]{mawet14}
{Mawet}, D., {Milli}, J., {Wahhaj}, Z., {et~al.} 2014, ArXiv e-prints

\bibitem[{{Mayer} {et~al.}(2005){Mayer}, {Wadsley}, {Quinn}, \&
  {Stadel}}]{2005MNRAS.363..641M}
{Mayer}, L., {Wadsley}, J., {Quinn}, T., \& {Stadel}, J. 2005, \mnras, 363, 641

\bibitem[{{Meschiari}(2012)}]{2012ApJ...761L...7M}
{Meschiari}, S. 2012, \apjl, 761, L7

\bibitem[{{Nelson}(2003)}]{2003MNRAS.345..233N}
{Nelson}, R.~P. 2003, \mnras, 345, 233

\bibitem[{{Neuhauser} \& {Brandner}(1998)}]{1998A&A...330L..29N}
{Neuhauser}, R. \& {Brandner}, W. 1998, \aap, 330, L29

\bibitem[{{Neuh{\"a}user} {et~al.}(2003){Neuh{\"a}user}, {Guenther}, {Alves},
  {Hu{\'e}lamo}, {Ott}, \& {Eckart}}]{neuhaeuser03}
{Neuh{\"a}user}, R., {Guenther}, E.~W., {Alves}, J., {et~al.} 2003,
  Astronomische Nachrichten, 324, 535

\bibitem[{{Nielsen} \& {Close}(2010)}]{2010ApJ...717..878N}
{Nielsen}, E.~L. \& {Close}, L.~M. 2010, \apj, 717, 878

\bibitem[{{Nielsen} {et~al.}(2013){Nielsen}, {Liu}, {Wahhaj}, {Biller},
  {Hayward}, {Close}, {Males}, {Skemer}, {Chun}, {Ftaclas}, {Alencar},
  {Artymowicz}, {Boss}, {Clarke}, {de Gouveia Dal Pino}, {Gregorio-Hetem},
  {Hartung}, {Ida}, {Kuchner}, {Lin}, {Reid}, {Shkolnik}, {Tecza}, {Thatte}, \&
  {Toomey}}]{2013ApJ...776....4N}
{Nielsen}, E.~L., {Liu}, M.~C., {Wahhaj}, Z., {et~al.} 2013, \apj, 776, 4

\bibitem[{{Nordstr{\"o}m} {et~al.}(2004){Nordstr{\"o}m}, {Mayor}, {Andersen},
  {Holmberg}, {Pont}, {J{\o}rgensen}, {Olsen}, {Udry}, \&
  {Mowlavi}}]{2004AA...418..989N}
{Nordstr{\"o}m}, B., {Mayor}, M., {Andersen}, J., {et~al.} 2004, A\&A, 418, 989

\bibitem[{{Orosz} {et~al.}(2012){Orosz}, {Welsh}, {Carter}, {Fabrycky},
  {Cochran}, {Endl}, {Ford}, {Haghighipour}, {MacQueen}, {Mazeh},
  {Sanchis-Ojeda}, {Short}, {Torres}, {Agol}, {Buchhave}, {Doyle}, {Isaacson},
  {Lissauer}, {Marcy}, {Shporer}, {Windmiller}, {Barclay}, {Boss}, {Clarke},
  {Fortney}, {Geary}, {Holman}, {Huber}, {Jenkins}, {Kinemuchi}, {Kruse},
  {Ragozzine}, {Sasselov}, {Still}, {Tenenbaum}, {Uddin}, {Winn}, {Koch}, \&
  {Borucki}}]{2012Sci...337.1511O}
{Orosz}, J.~A., {Welsh}, W.~F., {Carter}, J.~A., {et~al.} 2012, Science, 337,
  1511

\bibitem[{{Pallavicini} {et~al.}(1992){Pallavicini}, {Randich}, \&
  {Giampapa}}]{1992A&A...253..185P}
{Pallavicini}, R., {Randich}, S., \& {Giampapa}, M.~S. 1992, \aap, 253, 185

\bibitem[{{Pascucci} {et~al.}(2008){Pascucci}, {Apai}, {Hardegree-Ullman},
  {Kim}, {Meyer}, \& {Bouwman}}]{2008ApJ...673..477P}
{Pascucci}, I., {Apai}, D., {Hardegree-Ullman}, E.~E., {et~al.} 2008, \apj,
  673, 477

\bibitem[{{Perets}(2010)}]{2010arXiv1001.0581P}
{Perets}, H.~B. 2010, ArXiv e-prints

\bibitem[{{Pierens} \& {Nelson}(2008)}]{pierens08}
{Pierens}, A. \& {Nelson}, R.~P. 2008, \aap, 483, 633

\bibitem[{{Pierens} \& {Nelson}(2013)}]{pierens13}
{Pierens}, A. \& {Nelson}, R.~P. 2013, \aap, 556, A134

\bibitem[{{Rafikov}(2013)}]{2013ApJ...764L..16R}
{Rafikov}, R.~R. 2013, \apjl, 764, L16

\bibitem[{{Raghavan} {et~al.}(2010){Raghavan}, {McAlister}, {Henry}, {Latham},
  {Marcy}, {Mason}, {Gies}, {White}, \& {ten Brummelaar}}]{2010ApJS..190....1R}
{Raghavan}, D., {McAlister}, H.~A., {Henry}, T.~J., {et~al.} 2010, \apjs, 190,
  1

\bibitem[{{Rameau} {et~al.}(2013{\natexlab{a}}){Rameau}, {Chauvin}, {Lagrange},
  {Klahr}, {Bonnefoy}, {Mordasini}, {Bonavita}, {Desidera}, {Dumas}, \&
  {Girard}}]{2013AA...553A..60R}
{Rameau}, J., {Chauvin}, G., {Lagrange}, A.-M., {et~al.} 2013{\natexlab{a}},
  A\&A, 553, A60

\bibitem[{{Rameau} {et~al.}(2013{\natexlab{b}}){Rameau}, {Chauvin}, {Lagrange},
  {Meshkat}, {Boccaletti}, {Quanz}, {Currie}, {Mawet}, {Girard}, {Bonnefoy}, \&
  {Kenworthy}}]{2013arXiv1310.7483R}
{Rameau}, J., {Chauvin}, G., {Lagrange}, A.-M., {et~al.} 2013{\natexlab{b}},
  ArXiv e-prints

\bibitem[{{Song} {et~al.}(2003){Song}, {Zuckerman}, \&
  {Bessell}}]{2003ApJ...599..342S}
{Song}, I., {Zuckerman}, B., \& {Bessell}, M.~S. 2003, ApJ, 599, 342

\bibitem[{{Surkova} \& {Svechnikov}(2004)}]{2004yCat.5115....0S}
{Surkova}, L.~P. \& {Svechnikov}, M.~A. 2004, VizieR Online Data Catalog, 5115,
  0

\bibitem[{{Tamura}(2009)}]{tamura09}
{Tamura}, M. 2009, in American Institute of Physics Conference Series, Vol.
  1158, American Institute of Physics Conference Series, ed. T.~{Usuda},
  M.~{Tamura}, \& M.~{Ishii}, 11--16

\bibitem[{{Thalmann} {et~al.}(2009){Thalmann}, {Carson}, {Janson}, {Goto},
  {McElwain}, {Egner}, {Feldt}, {Hashimoto}, {Hayano}, {Henning}, {Hodapp},
  {Kandori}, {Klahr}, {Kudo}, {Kusakabe}, {Mordasini}, {Morino}, {Suto},
  {Suzuki}, \& {Tamura}}]{2009ApJ...707L.123T}
{Thalmann}, C., {Carson}, J., {Janson}, M., {et~al.} 2009, \apjl, 707, L123

\bibitem[{{Thalmann} {et~al.}(2013){Thalmann}, {Desidera}, {Bergfors},
  {Boccaletti}, {Bonavita}, {Carson}, {Feldt}, {Goto}, {Henning}, {Janson},
  {Klahr}, {Marzari}, \& {Mordasini}}]{2013prpl.conf2K012T}
{Thalmann}, C., {Desidera}, S., {Bergfors}, C., {et~al.} 2013, in Protostars
  and Planets VI, Heidelberg, July 15-20, 2013. Poster \#2K012, 12

\bibitem[{{Thebault}(2011)}]{2011CeMDA.111...29T}
{Thebault}, P. 2011, Celestial Mechanics and Dynamical Astronomy, 111, 29

\bibitem[{{Tokovinin} {et~al.}(2012){Tokovinin}, {Hartung}, {Hayward}, \&
  {Makarov}}]{tokovinin12}
{Tokovinin}, A., {Hartung}, M., {Hayward}, T.~L., \& {Makarov}, V.~V. 2012,
  \aj, 144, 7

\bibitem[{{Tokovinin} {et~al.}(2014){Tokovinin}, {Mason}, \&
  {Hartkopf}}]{tokovinin14}
{Tokovinin}, A., {Mason}, B.~D., \& {Hartkopf}, W.~I. 2014, \aj, 147, 123

\bibitem[{{Torres} {et~al.}(2006){Torres}, {Quast}, {da Silva}, {de La Reza},
  {Melo}, \& {Sterzik}}]{2006A&A...460..695T}
{Torres}, C.~A.~O., {Quast}, G.~R., {da Silva}, L., {et~al.} 2006, \aap, 460,
  695

\bibitem[{{Torres} {et~al.}(2008){Torres}, {Quast}, {Melo}, \&
  {Sterzik}}]{2008hsf2.book..757T}
{Torres}, C.~A.~O., {Quast}, G.~R., {Melo}, C.~H.~F., \& {Sterzik}, M.~F. 2008,
  {Young Nearby Loose Associations}, ed. {Reipurth, B.}, 757

\bibitem[{{Torres} {et~al.}(2003){Torres}, {Guenther}, {Marschall},
  {Neuh{\"a}user}, {Latham}, \& {Stefanik}}]{2003AJ....125..825T}
{Torres}, G., {Guenther}, E.~W., {Marschall}, L.~A., {et~al.} 2003, ApJ, 125,
  825

\bibitem[{{Trilling} {et~al.}(2007){Trilling}, {Stansberry}, {Stapelfeldt},
  {Rieke}, {Su}, {Gray}, {Corbally}, {Bryden}, {Chen}, {Boden}, \&
  {Beichman}}]{2007ApJ...658.1289T}
{Trilling}, D.~E., {Stansberry}, J.~A., {Stapelfeldt}, K.~R., {et~al.} 2007,
  \apj, 658, 1289

\bibitem[{{Veras} \& {Armitage}(2004)}]{2004MNRAS.347..613V}
{Veras}, D. \& {Armitage}, P.~J. 2004, MNRAS, 347, 613

\bibitem[{{Vigan} {et~al.}(2012){Vigan}, {Patience}, {Marois}, {Bonavita}, {De
  Rosa}, {Macintosh}, {Song}, {Doyon}, {Zuckerman}, {Lafreni{\`e}re}, \&
  {Barman}}]{2012AA...544A...9V}
{Vigan}, A., {Patience}, J., {Marois}, C., {et~al.} 2012, A\&A, 544, A9

\bibitem[{{Wang} {et~al.}(2014){Wang}, {Xie}, {Barclay}, \& {Fischer}}]{wang14}
{Wang}, J., {Xie}, J.-W., {Barclay}, T., \& {Fischer}, D.~A. 2014, \apj, 783, 4

\bibitem[{{Washuettl} \& {Strassmeier}(2001)}]{2001A&A...370..218W}
{Washuettl}, A. \& {Strassmeier}, K.~G. 2001, \aap, 370, 218

\bibitem[{{Welsh} {et~al.}(2013){Welsh}, {Orosz}, {Carter}, \&
  {Fabrycky}}]{2013arXiv1308.6328W}
{Welsh}, W.~F., {Orosz}, J.~A., {Carter}, J.~A., \& {Fabrycky}, D.~C. 2013,
  ArXiv e-prints

\bibitem[{{Wonnacott} {et~al.}(1993){Wonnacott}, {Kellett}, \&
  {Stickland}}]{1993MNRAS.262..277W}
{Wonnacott}, D., {Kellett}, B.~J., \& {Stickland}, D.~J. 1993, \mnras, 262, 277

\bibitem[{{Wright} {et~al.}(2011){Wright}, {Fakhouri}, {Marcy}, {Han}, {Feng},
  {Johnson}, {Howard}, {Fischer}, {Valenti}, {Anderson}, \&
  {Piskunov}}]{2011PASP..123..412W}
{Wright}, J.~T., {Fakhouri}, O., {Marcy}, G.~W., {et~al.} 2011, \pasp, 123, 412

\bibitem[{{Wright} {et~al.}(2004){Wright}, {Marcy}, {Butler}, \&
  {Vogt}}]{wright04}
{Wright}, J.~T., {Marcy}, G.~W., {Butler}, R.~P., \& {Vogt}, S.~S. 2004, \apjs,
  152, 261

\bibitem[{{Zuckerman} {et~al.}(2011){Zuckerman}, {Rhee}, {Song}, \&
  {Bessell}}]{2011ApJ...732...61Z}
{Zuckerman}, B., {Rhee}, J.~H., {Song}, I., \& {Bessell}, M.~S. 2011, \apj,
  732, 61

\bibitem[{{Zuckerman} \& {Song}(2004)}]{2004ARAA..42..685Z}
{Zuckerman}, B. \& {Song}, I. 2004, ARA\&A, 42, 685

\end{thebibliography}

\clearpage

\begin{appendix}


\section{Discussion of individual targets}
\label{s:appendix}

{\bf HIP 9892 = HD 13183:} 
Member of Tucana association, classified as a long-period SB by 
\citet{2007astro.ph..1293G}. No orbital solution is available.


\medskip \noindent
{\bf HIP 12545 = BD +05 0378:}
Member of $\beta$ Pic moving group. 
Identifed as SB1 in \cite{2003ApJ...599..342S} (peak-to-valley variation of 20 km/s,
no orbital solution provided). However, \cite{2012ApJ...749...16B} found no evidence for large RV variations from their
monitoring over 600 days (14 epochs, scatter of 179 m/s). These observations might be explained by
a high-eccentricity orbit.

\medskip \noindent
{\bf  UX~For = HIP 12716 = HD 17084:}  
Triple system formed by a close SB2 \citep[orbital period 0.9548\,d, mass ratio 1.371;][]{2001A&A...370..218W}
and an outer component, resolved by \citet{2012AJ....143...42H} and our own NACO observations, which is 
likely responsible of the proper motion difference between Hipparcos and 
historical proper motions \citep{2005AJ....129.2420M}.
The large activity levels are likely due to tidal locking of the inner components, as the Lithium equivalent width (Li EW) 
is similar to that of the Hyades. The presence of the third component significantly limits the
parameter space of possible planets around the central pair. 


\medskip \noindent
{\bf  HIP 16853 = HD  22705:}
Member of Tucana association. 
Astrometric orbit and parallax adopted from \citet{2007ApJ...654L..81M}.

\medskip \noindent
{\bf V1136 Tau = CHR~14 = HD 284163 = HIP 19591:}
Triple system, member of the Hyades.
A 2.39\,d RV orbit has been derived by \citet{1981AJ.....86..588G}, with indication
of the presence of a tertiary companion from the presence in the spectra of
an additional system of lines at constant radial velocity.
The inner pair was resolved as SB2 thanks to NIR RVs by \citet{2008ApJ...689..416B}, yielding 
a mass ratio of $0.68\pm0.03$.
The third component was visually resolved allowing a preliminary estimate of the orbit
\citep{2012A&A...546A..69M}.
The tertiary component is also identified in our NACO images (see Section~\ref{s:obs}).


\medskip \noindent
{\bf TYC 5907-1244-1 =  BD-20 951:}
Member of Tucana associaton and SB2 following \citet{2008hsf2.book..757T}.
No orbital solution is available. 

\medskip \noindent
{\bf AF Lep = HIP 25486 = HD 35850:} Member of $\beta$ Pic moving group. 
Reported as SB2 in \cite{2004AA...418..989N}, with mass ratio 
between the components of 0.72. Details of orbital solution not provided.

\medskip \noindent
{\bf HIP 25709 = HD 36329:} Member of Columba association, was 
reported as a SB2 with similar 
components by \citet{2008hsf2.book..757T}. The orbital solution is not 
available. 

\medskip \noindent
{\bf XZ~Pic = HIP 27134 =  CD-59 1125:}
Short-period SB1 according to sparse RV measurements 
\citep{2007A&A...470.1201D,2006A&A...460..695T,1999A&AS..138...87C}.
Age indicators based on activity and rotation are likely biased by tidal 
locking between the components.
The lithium in the spectrum indicates an age of about 300 Myr.



\medskip \noindent
{\bf 26~Gem = HIP 32104 = HD 48097 = HR 2466:}
Member of Columba association according to \citet{2011ApJ...732...61Z} and  \citet{2013ApJ...762...88M}.
Spectroscopic \citep{2005AA...443..337G} and astrometric (Hipparcos orbital solution) binary.
Combining the spectroscopic solution with the inclination from Hipparcos results
in a companion mass of $0.51 M_{\odot}$ at 1.87\,AU.
The secondary is most likely responsbile for the X-ray emission from the system.


\medskip \noindent
{\bf Alhena = $\boldsymbol{\gamma}$ Gem = HIP 31681 = HD 47105:}
Spectroscopic, astrometric and close visual binary. The constraints on orbital
solution derived from our own observations are presented in Section~\ref{s:binastro}.
The isochrone fitting yields an age of 300 Myr and the X-ray emission
(assuming it comes from the late-type secondary) is consistent with such 
an estimate. The space velocities are close to those of the Ursa Major 
association.

\medskip \noindent
{\bf TYC 8104-0991-1 = CD-45 2654:}
SB3 discovered by \citet{2006A&A...460..695T}, but no additional information
is available on the binary parameters.
The large Li EW indidates an age of about 100 Myr. 

\medskip \noindent
{\bf EM Cha = RECX7:}
SB2, member of $\eta$ Cha open cluster.
Masses and orbital parameters from \citet{2003MNRAS.338..616L}
(eccentricity not provided, we assume 0.0 due the short period).
The orbital period is equal to the photometric period, suggesting 
the occurrence of tidal locking.

\medskip \noindent
{\bf TYC 8569-3597-1 =  CD-53 2515:}
Member of Carina MG and SB2 with a period of 24.06 days \citep{2008hsf2.book..757T}.
Distance from \citet{2008hsf2.book..757T}.

\medskip \noindent
{\bf GS~Leo = HIP 46637 = HD 82159:}
Triple system, formed by a close pair and an outer component. The latter was observed
by \cite{2014arXiv1405.1560C}.
Stellar and binary parameters from \cite{2014arXiv1405.1559D}.

\medskip \noindent
{\bf TYC 9399-2452-1  =  HD 81485B:}
The SB2 HD 81485B \citep{1997A&A...328..187C} is part of a quadruple system with 
HD 81485A and its close visual companion at 9\arcsec{} $\sim$ 500 AU projected separation 
at the trigonometric distance of HD 81485A.
The SB2 component shows exceptional high levels of chromospheric activity, while the primary is 
only moderately active \citep{2006A&A...454..553D}. This, together with the Li EW comparable 
with that of the Hyades \citep{2006A&A...460..695T}, indicates 
an age of about 800 Myr and probable tidal-locking of the SB2 component.
The very young age for the primary (14 Myr) derived from isochrone fitting
by \cite{1998A&A...330L..29N} is probably due to
the unrecognized (at that time) multiplicity of A, that is actually
overluminous by about 0.5 mag in $K$-band with respect to the individual
Aa component.
Finally, we note discrepant trigonometric parallaxes between the two reductions of 
Hipparcos data.


\medskip \noindent
{\bf HIP 47760 = HD 84323:}    
Star flagged as SB1 in \citet{2002A&A...384..491C} without further details.
No orbital solution is available.
The Li EW larger than Pleiades of similar colors and similar to that of member of Tucana 
association but the lack of system RV prevents a proper evaluation of MG membership. 
The BANYAN II online tool \citep{2014ApJ...783..121G} without using RV information 
does not support membership in any of the MG considered in that work. We adopt an age of 30 Myr.

\medskip \noindent
{\bf TYC 6604-0118-1 = CD-22 7788 = BD-21 2961:}
Short-period (1.83\,d) SB2 \citep{2003AJ....125..825T}.
The Li EW indicate an age pf about 100\,Myr and the kinematic parameters
are compatible with such a moderately young age.
The close pair has a wide common proper motion companion (2MASS J09590930-2239582) 
at 26\arcsec{} projected separation.

\medskip \noindent
{\bf HS~Lup = HIP 74049 = HD 133822:}   
SB2 with period 17.83 d \citep{1961RGOB...30...93E}.
There is independent evidence that the star is rather young, as it is active with a possible 
rotation period which is different from the orbital one \citep{1989IBVS.3306....1H} and shows lithium 
intermediate between that of members of Hyades and Pleiadesof similar color.
We adopt an age of 250\,Myr.


\medskip \noindent
{\bf HIP 76629 = HD 139084 = V343 Nor:}
Triple system, member of $\beta$ Pic MG.
Evidence of RV variability is derived in the literature from \citet{2007astro.ph..1293G} and
\citet{2006A&A...460..695T}.
Including RVs measured on archival FEROS and HARPS spectra, we derived a tentative orbital solution
with a period of about 4.5 years and moderate eccentricity (0.5--0.6). The corresponding minimum mass of the companion is $0.11\,M_{\odot}$.
There is an additional companion, the M5Ve star HD 139084B at 10\arcsec{}.


\medskip \noindent
{\bf HIP 78416 = HD 143215:}
Triple system, formed by an SB2 system discovered by \citet{2006A&A...454..553D}
and a wide companion at 6\farcs55 (550\,AU projected separation).
The system is formed by three stars with similar masses.
The large X-ray emission might be due to fast rotation due to tidal circularization.
However, the isolated secondary also shows a fast rotational velocity, 
Ca II HK emission level similar to Pleiades, and
a Lithium 6708\,\AA\ line stronger than Pleiades stars of similar color.
Li lines are clearly visible and well separated also in the SB2 component.
The star is at a distance of $84\pm12$\,pc in front of the Lupus cloud.
The kinematical parameters and the space position are compatible with those
of Upper Centaurus Lupus (UCL) group. The age resulting from lithium is 
also fully compatible with UCL membership.




%


\medskip \noindent
{\bf BS~Ind = HIP 105404 = HD 202917:}
This is a spectroscopic binary with $P=3.3$\,yr and $e=0.6$, in which the primary is 
found to be itself an 
eclipsing binary with $P=0.43$ days. The components of the eclipsing system are 
likely late K- or early M-type stars, 
with a total mass of about 0.9 $M_{\odot}$, while the wider spectroscopic companion is a 
K0V-type star of $0.8 M_{\odot}$ \citep[see][for all the details]{2005AA...433..629G}.
Membership to Tuc-Hor association was supported by \cite{2004ARAA..42..685Z} and 
\citet{2013ApJ...762...88M} but rejected by \cite{2008hsf2.book..757T}.
The Li EW suggests an age as young as 10\,Myr but should be taken with caution because of the
composite nature of the system.
We adopt Tuc-Hor membership and its age.

\medskip \noindent
{\bf IK Peg = HIP 105860 = HD 204188:}
Single-lined SB with a massive white dwarf companion \citep{1993MNRAS.262..277W}.
The system should have evolved through the common envelope phase, rendering it
challenging to
estimate the original configuration and the age of the system \citep{2010MNRAS.403..179D}.
The position of the primary close to the zero-age main sequence, the large mass and hot 
temperature
of the white dwarf, and the young-disk kinematics support a moderately young age.
\citet{2007ApJ...658.1289T} adopt an age of\,100 Myr. 
The primary is also a $\delta$~Scu pulsating variable.

\medskip \noindent
{\bf $\boldsymbol{\delta}$ Cap = HIP 107556 =  HD 207098 = GJ 837:}
Short-period spectroscopic and semi-detached eclipsing binary.
Orbit from SB9: $P=1.02$ d, $e=0.01$, $K$1 = 70.80\,km/s.
In the catalog of semi-detached binaries \citep{2004yCat.5115....0S} individual masses 
and radii $M_{A}=1.50$, $M_{B}=0.56$, $R_{A}=1.85$, $R_{B}=1.56$ are listed.
The large X-ray luminosity of the system is likely due to tidal-spin up of 
the late type secondary.
An age of 540\,Myr is derived by \citet{2013ApJ...776....4N}. However, it is
highly uncertain. Furthermore, the position on CMD and then age from iscochrone 
may have been  altered by mass trasfer process.
The kinematic parameters are compatible with a young age.

\medskip \noindent
{\bf CS~Gru = HIP 109901 =  HD 211087:}
Lithium-rich K-type dwarf identified as SB1 (RMS of RV 37\,km/s, 4 measurements) in SACY.
The possibility of a tidally-locked binary can not be excluded from current data, but
the large lithium EW indicates an age of about 100\,Myr. 
The large uncertainty in the kinematic parameters due to the RV variability prevents a proper
evaluation of membership to associations.


\medskip \noindent
{\bf TYC 6386-0896-1 = HD 215341:} 
SB2 discovered by \citet{2002IBVS.5281....1C}; the orbital solution is not available.
The very large coronal and chromospheric activity are likely due to tidal locking, while
Lithium is within the locus of Pleides stars. We adopt an age of 150\,Myr.



\end{appendix}

\end{document}